\documentclass[aps,amssymb,noshowpacs,11pt,eqsecnum,superscriptaddress,nofootinbib]{revtex4}

\usepackage{graphicx,graphics}
\usepackage{enumerate}
\usepackage{amsfonts}
\usepackage{amssymb}

\usepackage{xcolor}
\usepackage{colortbl}
\usepackage{pstricks,multido}
\usepackage{wrapfig}

\definecolor{lightyellow}{rgb}{.95,.96,.5}
\definecolor{myblue}{rgb}{.39,.54,0.82}
\definecolor{midblue}{rgb}{.68,.68,1}
\definecolor{lightblue}{rgb}{.91,.91,1}
\definecolor{mygrey}{rgb}{.75,.75,.75}
\definecolor{lightred}{rgb}{1.,0.33,0.09}
\definecolor{mybrown}{rgb}{0.69,0.49,0.30}
\definecolor{lightlightblue}{rgb}{.85,.85,1}
\definecolor{midlightblue}{rgb}{.8,.8,1}

\usepackage{enumitem}

\catcode`\@=11
\def\numberbysection{\@addtoreset{equation}{section}
     \def\theequation{\arabic{section}.\arabic{equation}}}
\numberbysection

\def\be{\begin{equation}}
\def\ee{\end{equation}}
\newcommand\bea{\begin{eqnarray}}
\newcommand\eea{\end{eqnarray}}
\renewcommand\phi{\varphi}

\def\benn{\begin{eqnarray*}}
\def\eenn{\end{eqnarray*}}
\def\half{{\textstyle {1 \over 2}}}

\def\ra{\rangle}
\def\Z{{\mathbb Z}}

\def\C{{\mathbb C}}

\def\half{{\textstyle {1\over 2}}}

\def\ci{{\rm i}}
\def\a{\alpha}

\def\ep{\varepsilon}
\def\s{\sigma}
\def\bi{{\mathbb I}}

\def\hdimer{\pspolygon[linearc=.35,linecolor=black,fillstyle=solid,fillcolor=lightgray](-0.35,-0.35)(-0.35,0.35)(1.35,0.35)(1.35,-0.35)}
\def\vdimer{\pspolygon[linearc=.35,linecolor=black,fillstyle=solid,fillcolor=mybrown](-0.35,-0.35)(-0.35,1.35)(0.35,1.35)(0.35,-0.35)}

\def\aup{\psline[linewidth=1.5pt,linecolor=mybrown]{->}(0,0.15)(0,0.85)}
\def\ado{\psline[linewidth=1.5pt,linecolor=lightgray]{<-}(0,0.15)(0,0.85)}
\newcommand{\V}{\mathcal{V}}
\newcommand{\ch}{\mathrm{ch}}

\def\vvdots{\mathinner{\mkern1mu\raise1pt\vbox{\kern7pt\hbox{.}}\mkern2mu
  \raise4pt\hbox{.}\mkern2mu\raise7pt\hbox{.}\mkern1mu}}

\newcommand{\ri}{{\rm i}}

\begin{document}

\title{Refined conformal spectra in the dimer model}

\author{J{\o}rgen Rasmussen}
\email{j.rasmussen@uq.edu.au}
\affiliation{School of Mathematics and Physics, 
University of Queensland, St Lucia, Brisbane, Queensland 4072, Australia}

\author{Philippe Ruelle}
\email{philippe.ruelle@uclouvain.be}
\affiliation{Institute for Research in Mathematics and Physics, 
Universit\'{e} catholique de Louvain, B-1348 Louvain-la-Neuve, Belgium}

\begin{abstract}
Working with Lieb's transfer matrix for the dimer model, we point out that the full set of dimer configurations may be partitioned into disjoint subsets (sectors) closed under the action of the transfer matrix. These sectors are labelled by an integer or half-integer quantum number we call the variation index. In the continuum scaling limit, each sector gives rise to a representation of the Virasoro algebra. We determine the corresponding conformal partition functions and their finitizations, and observe an intriguing link to the Ramond and Neveu-Schwarz sectors of the critical dense polymer model as described by a conformal field theory with central charge $c=-2$.
\end{abstract}

\maketitle

\section{Introduction}

In his 1967 paper~\cite{lieb}, Lieb proposed a transfer matrix for the dimer model~\cite{FR37,Kasteleyn61,TF61,Fisher61}. It is our impression that this work has not been fully exploited and that it may shed light on the status of the dimer model as a conformal field theory (CFT). A certain confusion has prevailed for some time regarding the central charge of the relevant CFT. Indeed, some aspects of the dimer model can be described~\cite{iprh} by a CFT with $c=-2$~\cite{Flohr96,Kausch9510,GK96,gaka,pera,PRV}, hinting at its logarithmic character~\cite{Gurarie93,Flohr03,Gab03,PRZ}, while some other aspects, most notably those involving monomers~\cite{aupe,fms,plf}, seem to require~\cite{pr,dub} a description in terms of a CFT with $c=1$. Moreover, in certain geometries, the dimer model is equivalent~\cite{Temperley74,MD92} to the Abelian sandpile model~\cite{BTW88,Dhar90,Dhar99}, which is widely believed~\cite{maru,ru1,jpr,ru2,ppr} to be described by a logarithmic CFT with $c=-2$.

Here we consider dimers defined on rectangular lattices with or without periodic boundary conditions on the horizontal or vertical edges. Working with Lieb's transfer matrix, we find that the full set of dimer configurations may be partitioned into disjoint subsets we call {\em sectors}. These sectors are closed under the action of the transfer matrix and are naturally labelled by an integer or half-integer quantum number we call the {\em variation index}. In Lieb's spin picture, this quantum number is obtained as the eigenvalue of a particular matrix defined in terms of Pauli matrices.

In the case with open boundary conditions, each sector is found to be associated with a representation of the Virasoro algebra as the corresponding sector partition function yields a Virasoro character in the continuum scaling limit. These characters can be organised in a manner consistent with a CFT with central charge $c=-2$. Furthermore, the corresponding finitizations are accounted for by a combinatorial enumeration reminiscent of the physical combinatorics underlying the critical dense polymer model on the strip~\cite{pera} and cylinder~\cite{PRV}. This gives rise to a separation into Ramond and Neveu-Schwarz sectors. Following~\cite{Sal92,PRV}, we are thus using the terminology of supersymmetry even though we do not claim any superconformal symmetry in the dimer model.

In the case of even system sizes with periodic boundary conditions, each sector can be described as composed of two `chiral halves', much akin to the critical dense polymer model on the cylinder~\cite{PRV} and many other similar lattice models. 
The two `chiral halves' are not independent, but matched up by imposing a simple gluing condition. The combinatorial description of the ensuing finitized partition functions again resemble the physical combinatorics of the critical dense polymer model. 

Despite the obvious parallels between our treatment of the conformal properties of the dimer model and the similar studies of the critical dense polymer model~\cite{pera,PRV}, there are crucial differences. In particular, the variation index is a {\em good} quantum number, while the related so-called defect number in~\cite{pera,PRV} is not. In addition, the variation index takes on negative as well as positive values, again unlike the defect number which is non-negative. These differences ensure that the physical partition function for the dimer model on a torus with both lattice sides even is {\em modular invariant}, as opposed to the toroidal partition functions discussed in~\cite{PRV}. This modular invariant partition function for dimers was originally obtained by Ferdinand~\cite{ferd}, and has since appeared as the partition function for symplectic fermions as described by the so-called triplet model~\cite{gaka}.

The layout of this paper is as follows. In Section~\ref{SecDimers}, we review the basic formulation of the transfer matrix, as proposed by Lieb \cite{lieb}. Section~\ref{SecSectors} examines the partitioning of the configuration space into an (in the continuum scaling limit) infinite number of sectors left invariant by the transfer matrix $T$. These sectors correspond to orbits under $T$ and are thus the finest possible partitioning into invariant spaces. The sectors are interpreted in terms of eigenspaces of a certain operator $\V$, whose main property is to commute with the squared transfer matrix $T^2$. Section~\ref{Section:ConformalSpectra} generalizes Lieb's diagonalisation of the (squared) transfer matrix to the open boundary conditions, and works out the relevant partition functions. Sections~\ref{Section:SectorsSpectra} and \ref{SecCFT} extend this to the individual sectors, whose conformal contents are discussed. This is made more explicit in Section~\ref{SecOpen} and \ref{SecPeriodic}, respectively for the open and periodic boundary conditions, where the combinatorics of the conformal states is related to that in the critical dense polymer model \cite{pera,PRV}. Concluding remarks are presented in Section~\ref{SecConclusion}. Jacobi's theta functions are reviewed in
Appendix~\ref{AppTheta}, while Appendix~\ref{AppBFE} contains some technical details.
\newpage
\noindent
{\bf Note added.}
\\
After completion of the present work, Jesper L.$\!\,$ Jacobsen has informed us that the existence of sectors in the dimer model had been observed before, albeit not in the framework of Lieb's transfer matrix, nor applied to the computation of conformal partition functions. Two descriptions have been proposed. In the first one, the separation into sectors is based on the geometric observation that the superposition of a general dimer configuration and a certain fixed configuration results in the formation of a number of `strings'. In this view, a sector collects all dimer configurations giving rise to the same number of strings. The details of this construction can be found in~\cite{AIJMP,jacob}. The other, more algebraic approach has been discussed in the context of trimer tilings~\cite{GDJ}, but works more generally for $p$-mers, $p \geq 2$. In this second description, a certain function on $p$ letters is associated with a `profile', namely a line that goes through the lattice from left to right without crossing any of the $p$-mers. This function is constant on the set of all profiles which can be obtained from each other by adding $p$-mers (i.e. the portion of the lattice in-between two profiles can be covered by $p$-mers) and defines the sectors. For $p=2$, this function essentially coincides with our variation index. Despite these earlier observations, we believe that our applications of Lieb's transfer matrix provide new insight into the sector structure as well as the dimer model more generally.

\section{Dimers and Lieb's transfer matrix}
\label{SecDimers}

\subsection{Dimer model}

Let us start by briefly recalling what the dimer model is. We consider a rectangular grid in $\Z^2$ with $M$ rows and $N$ columns (possibly with periodic boundary conditions in the vertical and/or horizontal directions), which we cover with dimers. A dimer is a rectangular tile ($1 \times 2$ or $2 \times 1$), like a domino, covering exactly two adjacent sites\footnote{There is a well-known generalisation in which monomers, which cover only one site each, are allowed, but we do not consider this case here.}. A dimer configuration is then an arrangement of dimers such that every site of the grid is covered by {\em exactly} one tile, thereby disallowing dimers to overlap. A primary objective in this model is to study the statistics of the dimer configurations. 

The configurations are weighted according to their tile content. Without loss of generality, we may decide that a vertical dimer has weight 1, while a horizontal dimer has weight $\a$, so that a generic configuration receives a weight equal to $\a^h$, with $h$ the number of horizontal dimers. The partition function sums the weights of all configurations,
\be
 Z_{M,N} = \sum_{\rm configs} \; \a^h .
\ee

%
\begin{figure}[b]
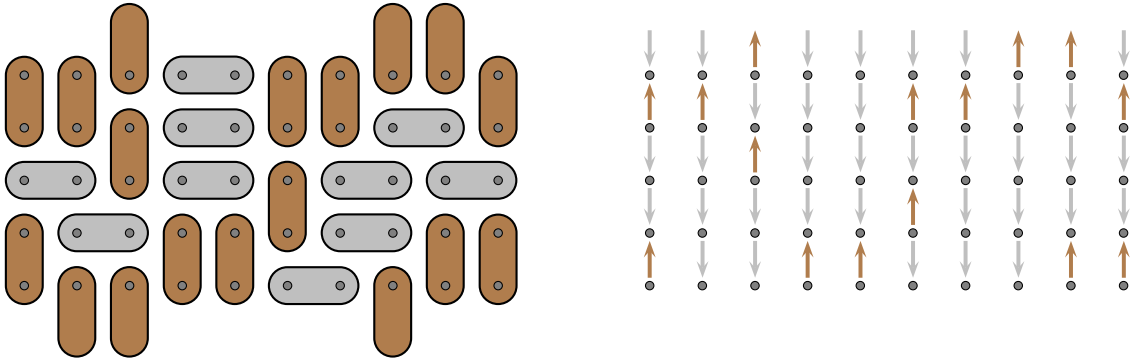

\psset{xunit=0.7cm}
\psset{yunit=0.7cm}
\psset{runit=0.7cm}
\pspicture(-1,-1)(10,6)
\rput(0,0){\vdimer}
\rput(1,-1){\vdimer}
\rput(2,-1){\vdimer}
\rput(3,0){\vdimer}
\rput(4,0){\vdimer}
\rput(5,0){\hdimer}
\rput(7,-1){\vdimer}
\rput(8,0){\vdimer}
\rput(9,0){\vdimer}
\rput(1,1){\hdimer}
\rput(5,1){\vdimer}
\rput(6,1){\hdimer}
\rput(0,2){\hdimer}
\rput(2,2){\vdimer}
\rput(3,2){\hdimer}
\rput(6,2){\hdimer}
\rput(8,2){\hdimer}
\rput(0,3){\vdimer}
\rput(1,3){\vdimer}
\rput(3,3){\hdimer}
\rput(5,3){\vdimer}
\rput(6,3){\vdimer}
\rput(7,3){\hdimer}
\rput(9,3){\vdimer}
\rput(2,4){\vdimer}
\rput(3,4){\hdimer}
\rput(7,4){\vdimer}
\rput(8,4){\vdimer}
\multido{\nt=0+1}{10}{\pscircle[linewidth=0.4pt,fillstyle=solid,fillcolor=gray](\nt,0){0.09}}
\multido{\nt=0+1}{10}{\pscircle[linewidth=0.4pt,fillstyle=solid,fillcolor=gray](\nt,1){0.09}}
\multido{\nt=0+1}{10}{\pscircle[linewidth=0.4pt,fillstyle=solid,fillcolor=gray](\nt,2){0.09}}
\multido{\nt=0+1}{10}{\pscircle[linewidth=0.4pt,fillstyle=solid,fillcolor=gray](\nt,3){0.09}}
\multido{\nt=0+1}{10}{\pscircle[linewidth=0.4pt,fillstyle=solid,fillcolor=gray](\nt,4){0.09}}
\endpspicture
\hspace{1.2cm}
\pspicture(0,-1)(10,6.5)
\rput(0,0){\aup}
\rput(1,0){\ado}
\rput(2,0){\ado}
\rput(3,0){\aup}
\rput(4,0){\aup}
\rput(5,0){\ado}
\rput(6,0){\ado}
\rput(7,0){\ado}
\rput(8,0){\aup}
\rput(9,0){\aup}
\rput(0,1){\ado}
\rput(1,1){\ado}
\rput(2,1){\ado}
\rput(3,1){\ado}
\rput(4,1){\ado}
\rput(5,1){\aup}
\rput(6,1){\ado}
\rput(7,1){\ado}
\rput(8,1){\ado}
\rput(9,1){\ado}
\rput(0,2){\ado}
\rput(1,2){\ado}
\rput(2,2){\aup}
\rput(3,2){\ado}
\rput(4,2){\ado}
\rput(5,2){\ado}
\rput(6,2){\ado}
\rput(7,2){\ado}
\rput(8,2){\ado}
\rput(9,2){\ado}
\rput(0,3){\aup}
\rput(1,3){\aup}
\rput(2,3){\ado}
\rput(3,3){\ado}
\rput(4,3){\ado}
\rput(5,3){\aup}
\rput(6,3){\aup}
\rput(7,3){\ado}
\rput(8,3){\ado}
\rput(9,3){\aup}
\rput(0,4){\ado}
\rput(1,4){\ado}
\rput(2,4){\aup}
\rput(3,4){\ado}
\rput(4,4){\ado}
\rput(5,4){\ado}
\rput(6,4){\ado}
\rput(7,4){\aup}
\rput(8,4){\aup}
\rput(9,4){\ado}
\multido{\nt=0+1}{10}{\pscircle[linewidth=0.4pt,fillstyle=solid,fillcolor=gray](\nt,0){0.09}}
\multido{\nt=0+1}{10}{\pscircle[linewidth=0.4pt,fillstyle=solid,fillcolor=gray](\nt,1){0.09}}
\multido{\nt=0+1}{10}{\pscircle[linewidth=0.4pt,fillstyle=solid,fillcolor=gray](\nt,2){0.09}}
\multido{\nt=0+1}{10}{\pscircle[linewidth=0.4pt,fillstyle=solid,fillcolor=gray](\nt,3){0.09}}
\multido{\nt=0+1}{10}{\pscircle[linewidth=0.4pt,fillstyle=solid,fillcolor=gray](\nt,4){0.09}}
\endpspicture
\caption{An arrow configuration is associated with a dimer configuration: a vertical dimer at a site $i$ becomes an up arrow (in brown) placed above $i$, while the absence of a vertical dimer is indicated by a down arrow (in grey).}
\end{figure}

\subsection{Transfer matrix}

In~\cite{lieb}, Lieb introduced a transfer matrix for this model, which builds all possible dimer configurations row by row with the direction of transfer upward. First, one replaces a dimer configuration by a set of arrows, up or down, attached to the sites, and placed above them, as illustrated in Figure 1. An up arrow at site $i$ means that there is a vertical dimer covering $i$ and its northern neighbour $i+\hat e_y$; a down arrow indicates the absence of such a vertical dimer. In the latter case, it either means, if the down arrow is right above an up arrow, that site $i$ pairs up with its neighbour below to form a vertical dimer; otherwise, it means that the down arrow at $i$ associates itself with another down arrow at the left or right neighbouring site, to form a horizontal dimer. This correspondence is indicated in Figure 1 for a tile configuration covering five successive rows. A row of $N$ sites can have $2^N$ different arrow configurations. It is natural to think of the up and down arrows as the canonical base elements $(1,0)$ and $(0,1)$ of a vector space $\C^2$. A row configuration of arrows can then be viewed as an element of the multiple tensor product $(\C^2)^{\otimes N}$. 

The next task is to define the operators which produce dimers; vertical or horizontal. Their action will be given in terms of tensor products of Pauli matrices, 
\be
 \s_i = {\mathbb I} \otimes \ldots \otimes {\mathbb I}\otimes \s \otimes{\mathbb I}\otimes \ldots \otimes {\mathbb I}, 
\ee
where $\s$, in position $i$, is one of the following three matrices (given in the basis $|\!\uparrow\rangle,\, |\!\downarrow\rangle$)
\be
 \s^x = \pmatrix{0 & 1 \cr 1 & 0}, \qquad \s^- = \pmatrix{0 & 0 \cr 1 & 0}, \qquad\s^+ = \pmatrix{0 & 1 \cr 0 & 0}.
\ee
The matrices on different sites commute, $\s_i\s_j=\s_j\s_i$ for all $i \neq j$, while those referring to the same site satisfy $(\s_i^-)^2 = (\s_i^+)^2 = (\s_i^x)^2-{\mathbb I} = 0$ and $\s_i^\pm \, \s_i^\mp \, \s_i^\pm = \s_i^\pm$ and hence $\s_i^x \, \s_i^\mp \, \s_i^x = \s_i^\pm$.

The transfer matrix acts on a horizontal sequence of arrows in two steps.
\begin{enumerate}
\item It first reverses all incoming arrows, by acting with 
\be
 V_1 = \prod_i \: \s_i^x.
\label{V1}
\ee 
This ensures that an up arrow becomes a down arrow, in accordance with the fact that an up arrow cannot propagate since two vertical dimers would then overlap. Similarly, a down arrow automatically propagates upward to an up arrow, corresponding to the lower half of a vertical dimer. 

\item After step 1, an outgoing up arrow at site $i$, corresponding to a vertical dimer, may be kept or changed into half a horizontal dimer. A horizontal dimer can be produced by reversing a {\em pair} of neighbouring up arrows, i.e.$\!\,$ by acting with $\s_i^-\s_{i+1}^-$, multiplied by the corresponding weight $\a$. The combination ${\mathbb I} + \a\, \s_i^-\s_{i+1}^-$ does not do anything or implements the change, so the operator 
\be
 V_3 \equiv \prod_i \: [{\mathbb I} + \a\, \s_i^-\s_{i+1}^-] = \exp{(\a \sum_i\,\s_i^-\s_{i+1}^-)}
\label{V3}
\ee 
offers this alternative for every pair of adjacent sites in a row.
\end{enumerate}

The product $V_3 V_1$ incorporates all possible ways to go from a row of specified arrows to the next one up while respecting the constraints: it creates horizontal dimers wherever possible, and correctly handles the incoming vertical dimers. It is the transfer matrix defined in~\cite{lieb}\footnote{Lieb's original matrix also allows for monomers in which case a third operator is needed in addition to $V_1$ and $V_3$. In his paper~\cite{lieb}, this operator is denoted by $V_2$.}
\be
 T = V_3 V_1 = \exp{(\sum_{i\geq 1} \, \a\,\s_i^-\s_{i+1}^-)} \, \prod_{j=1}^N \: \s_j^x.
\label{TV3V1}
\ee
If no horizontal periodicity is assumed, the summation over $i$ in the exponential runs from 1 to $N-1$, whereas in the periodic case, it runs to $N$, with the identification $\s_{N+1} = \s_1$.

The action of $T$ on an incoming row of arrows $|{\rm in}\rangle$ produces a linear combination of outgoing rows of arrows, the set of which represents all possible row configurations atop $|{\rm in}\rangle$. The coefficients in this linear combination involve powers of $\a$ which count the number of horizontal dimers. Likewise, the power $T^m$ iteratively constructs all possible arrays of $m$ rows built on top of $|{\rm in}\rangle$. The top row in this process, $|{\rm out}\rangle = T^m |{\rm in}\rangle$, comprises all possible outgoing row configurations, $m$ layers higher than the incoming row $|{\rm in}\rangle$. The configuration in Figure 1 represents a possible realisation of the action of $T^4$ on the bottom row, taken as incoming state $|{\rm in}\rangle = |\!\!\uparrow \downarrow \downarrow \uparrow \uparrow \downarrow \downarrow \downarrow \uparrow \uparrow \rangle$. In this particular case ($N=10$ with no horizontal periodicity), there are $24\,203$ other possible realisations compatible with $|{\rm in}\rangle$.

We note that, although the mapping of a dimer configuration to an arrow configuration is well-defined and unambiguous, the inverse mapping is locally not so. The source of ambiguity resides in the pairs of adjacent down arrows, which, locally, may correspond to a horizontal dimer or to two vertical dimers coming from below. The ambiguity can, however, be lifted by looking at the neighbourhood of these pairs: a pair of down arrows atop a pair of up arrows unambiguously corresponds to two vertical dimers. Thus, by looking at larger and larger neighbourhoods, and eventually at the whole grid, the mapping from arrows to dimers becomes globally unambiguous for a prescribed set of boundary conditions. 

We further emphasise the important role played by the boundary conditions as they may forbid certain rows of arrows. For example, the bottom row of arrows in Figure 1 would not be allowed if that row had been a straight boundary: the three adjacent down arrows in the right half necessarily give rise to an arrangement of dimers where a dimer protrudes from the grid.

\subsection{Partition functions}

The partition functions for various boundary conditions can be computed in terms of $T$ in the usual way. If vertical periodicity is imposed, we have 
\be
 Z_{M,N} = {\rm Tr}\, T^M
\label{ZMN}
\ee 
for periodic or non-periodic (open) horizontal boundary conditions. In case no vertical periodicity is imposed, the grid is bordered by two horizontal edges. On these edges, the usual condition is to restrict the dimers not to protrude from the boundaries. In this situation, the partition function is given by 
\be
 Z_{M,N} = \sum_{|{\rm in} \rangle} \; \langle {\rm out} | T^{M-1} | {\rm in} \rangle,
\label{zopen}
\ee
where the sum is over all arrow configurations $|{\rm in}\rangle$ allowed on the lower boundary, namely those where down arrows come in adjacent pairs. The out-state is given by 
\be 
 |{\rm out}\rangle = |\!\!\downarrow \downarrow \cdots \downarrow \rangle
\ee
since the top row cannot contain vertical dimers sticking out of the grid. The partition function (\ref{zopen}) is thus given by the sum of certain elements of the row corresponding to the arrow configuration $|{\rm out}\rangle$ in the matrix $T^{M-1}$. An obvious advantage of using the transfer matrix is that it allows, through formulae such as (\ref{zopen}), to consider other, new  boundary conditions on the bottom and top edges by choosing other sets of in- and out-states, thereby allowing the dimers to protrude in a specified way.

By construction of the pure dimer model, the partition function $Z_{M,N}$ vanishes if both $M$ and $N$ are odd. For convenience, we shall therefore assume $M$ even in the following. As we are mainly interested in situations with vertical periodicity as in (\ref{ZMN}), our analysis is primarily based on $T^2$ instead of $T$ itself.

\section{Sectors and variation index}
\label{SecSectors}

The transfer matrix is typically not irreducible on the space of arrow configurations; it has many invariant subspaces. If such a subspace is spanned by a set of arrow configurations, we refer to it as a $T$-sector. Note that a generic eigenspace of $T$ is not a $T$-sector, since it is spanned by a set of linear combinations of arrow configurations where the number of participating arrow configurations (generally) exceeds the dimension of the eigenspace. Similarly, a $T^2$-invariant subspace spanned by a set of arrow configurations is called a $T^2$-sector.

The simplest example of a $T$-sector is the subspace spanned by the two alternating sequences, 
\be
 |\!\downarrow \uparrow \downarrow \uparrow \cdots \rangle\qquad {\rm and} \qquad 
 |\!\uparrow \downarrow \uparrow \downarrow \cdots \rangle,
\label{alt}
\ee 
where $N$ must be even in the case of periodic boundary conditions. (Periodic boundary conditions with $N$ odd are discussed below.) As the transfer matrix maps one of these two states onto the other, they clearly form a two-dimensional invariant subspace. Individually, the two configurations form one-dimensional $T^2$-sectors. The other sectors are larger and have dimensions which grow with $N$. 

In the following, we will determine the finest possible decomposition of the set of arrow configurations into $T$- or $T^2$-sectors, and work out the corresponding dimensions. This classification depends on the parity of $N$ and whether periodicity is imposed.

\subsection{Variation index}

Let us introduce the operator
\be
 \V = \sum_{i=1}^N \,(-1)^i \, [\s^+_i\s^-_i - {\textstyle {1 \over 2}}] = 
 {\textstyle \frac{1}{2}}\sum_{i=1}^N(-1)^i\sigma_i^z=
 \cases{
 \displaystyle{\sum_{i=1}^N \,(-1)^i \, \s^+_i\s^-_i} &\ \ if $N$ is even,\cr
 \noalign{\medskip}
 \displaystyle{{\textstyle{1 \over 2}} + \sum_{i=1}^N \,(-1)^i \, \s^+_i\s^-_i} &\ \ if $N$ is odd.}
\ee
Each of the constituent operators 
\be
(-1)^i \, [\s^+_i\s^-_i - \half] = {\textstyle{(-1)^i \over 2}} \s^z_i,\qquad \text{where}\quad \s^z = \pmatrix{1 & 0 \cr 0 & -1},
\ee 
takes the value $+\frac{1}{2}$ or $-\frac{1}{2}$ depending on the parity of $i$ and on whether the arrow at $i$ is up or down. $\V$ is thus diagonal in the arrow basis and has eigenvalues $v$ of the form
\be
 v\in\{-n,\,-n+1,\,\ldots,\,n\},\qquad n = \frac{N}{2}.
\label{v}
\ee
 We note that a pair of equal neighbouring arrows in positions $i$ and $i+1$, $1\leq i<N$, brings a zero contribution to $v$, while two opposite neighbouring arrows in similar positions contribute $+1$ or $-1$ to $v$. We may therefore interpret $v$ as an accumulative measure of the local changes in an arrow configuration; for this reason, we call it the {\em variation index}. This variation index $v$ is an integer for $N$ even, but a half-integer for $N$ odd.
 
In the case of periodic boundary conditions, the variation index depends on where the labelling $i$ of the arrows begins. For $N$ even, an odd shift in the labelling will change the sign, whereas an even shift has no effect on the variation index. For $N$ odd, on the other hand, the starting point significantly affects the value of the variation index. 
 
For each value of $v$ in (\ref{v}), let $E_v$ be the vector (sub-)space spanned by the arrow configurations having variation index equal to $v$. For instance, $E_n$ is generated by the single state $|\!\downarrow \uparrow \downarrow \uparrow \cdots \rangle$ discussed above (\ref{alt}), while $E_{-n}$ is generated by its flip. It follows that ${\rm dim}\, E_n = {\rm dim}\,E_{-n} = 1$. More generally, we have 
\be 
 {\rm dim}\, E_v = {\rm dim}\,E_{-v}\,,
\ee 
since a global flip of arrows bijectively maps one of these spaces onto the other. Because each site contributes $\pm {1 \over 2}$ to $v$, the generating function for the dimensions of $E_v$ is given by
\be
 \sum_{v=-n}^n \: ({\rm dim}\,E_v) \, t^v = \Big(\!\sqrt{t} + {1 \over \sqrt{t}}\Big)^N,
\label{sumv}
\ee
where the summation index $v$ increments in steps of $1$, from which it follows that 
\be
 {\rm dim}\,E_v = {N \choose n-v} .
\ee

\subsection{Sectors}

Our classification of the finest possible sector decompositions of the space of arrow configurations is conveniently described in terms of the variation index $v$ and the spaces $E_v$. First, we show that certain simple direct sums of these spaces constitute $T$- and $T^2$-invariant spaces. We subsequently argue that these sectors are indeed the smallest possible ones. Our analysis depends on the parity of $N$ and whether periodic boundary conditions are imposed.

We first examine how the transfer matrix $T=\exp{(\sum_i \, \a\,\s_i^-\s_{i+1}^-)} \, \prod_j \s_j^x$ acts on $E_v$. Because $V_1=\prod_j \s_j^x$ induces a global flip, it anti-commutes with $\V$. Moreover, each constituent operator $\s_i^-\s_{i+1}^-$ commutes with $\V$ unless $i=N$, $N$ is odd and periodic conditions are imposed horizontally. Except in this particular case, $\V$ and $T$ therefore {\em anti-commute}. It follows that $T$ changes the sign\footnote{Likewise, $\V$ changes the sign of the eigenvalues of $T$, which therefore come in pairs of opposite sign except in the sector $E_0$ on which $\V$ vanishes.} of the eigenvalues of $\V$, thus mapping $E_v$ onto $E_{-v}$,
\be
 T:\quad E_v\ \to\ E_{-v}.
\label{TEE}
\ee 
Consequently, the following spaces are left invariant:
\be
 T\mbox{-sectors\ :}\qquad E_0,\quad E_v \oplus E_{-v},\quad v > 0.
\label{TE}
\ee
It also follows that $T^2$ {\em commutes} with $\V$ and thus leaves each {\em individual} space $E_v$ invariant,
\be
 T^2\mbox{-sectors\ :}\qquad E_v.
\label{T2E}
\ee

The $T$-{\em orbit} (or $T^2$-{\em orbit}) of a given arrow configuration is introduced as the linear span of the set of arrow configurations appearing in the expansions of repeated applications of $T$ (or $T^2$) on the given arrow configuration. 
From the explicit form of $T$, an element in the $T$-orbit of $\cal C$ is a linear combination of elements of the form
\be
{\cal C}' = \ldots \Big[(\s^-_{j_1}\s^-_{j_1+1})\,(\s^-_{j_2}\s^-_{j_2+1}) \ldots (\s^-_{j_\ell}\s^-_{j_\ell+1})\,V_1\Big]\Big[(\s^-_{i_1}\s^-_{i_1+1})\,(\s^-_{i_2}\s^-_{i_2+1}) \ldots (\s^-_{i_k}\s^-_{i_k+1})\,V_1\Big]\, \cal C,
\label{orbit}
\ee
where each factor in square brackets represents the action of one $T$. Within a given such factor,  the position labels of the $\s^-$ matrices are all different. 

We note that being in a given orbit is an equivalence relation. To see this, we observe that if the previous relation (\ref{orbit}) involves $q$ factors of $T$, namely ${\cal C}' = [T_q] \ldots [T_2]\, [T_1] \, {\cal C} \neq 0 $, then we also have ${\cal C} = [T_1] \, [T_2] \ldots [T_q] \, {\cal C}'$ because each factor squares to 1 (unless it annihilates the configuration it is acting on). This shows that the relation is symmetric, from which transitivity readily follows. 

As the orbits provide the maximal decomposition into invariant subspaces, it is enough to show that the sectors $E_0$ and $E_v \oplus E_{-v}$ in (\ref{TE}) coincide with orbits under $T$. As indicated, the above proof that $\V$ anti-commutes with $T$ breaks down in the periodic case with $N$ odd. Indeed, we show below that in this case, there is a single orbit so that $T$ is transitive on the full space. We start with the open case as it simplifies the discussion of the periodic case. 

To show that every $T$-sector in (\ref{TE}) is a $T$-orbit, it suffices to demonstrate that a particular canonical arrow configuration in the sector is an element of the orbit of any arrow configuration in the sector. Such a canonical arrow configuration is conveniently chosen as
\be
 \mathcal{C}_0=|\!\downarrow\downarrow\cdots\downarrow\rangle\in E_0\,, \qquad \Big({\rm or}\quad |\!\uparrow\uparrow\cdots\uparrow\rangle\in E_0\Big)
\label{C0}
\ee
and, for $v>0$, as
\be
\mathcal{C}_v=|\underbrace{\downarrow\uparrow\downarrow\uparrow\cdots\downarrow\uparrow}_{2\lfloor v\rfloor}\underbrace{\downarrow\downarrow\cdots\downarrow}_{N-2\lfloor v\rfloor}\rangle\in E_v\qquad {\rm or}\qquad
 \mathcal{C}_{-v}=|\underbrace{\uparrow\downarrow\uparrow\downarrow\cdots\uparrow\downarrow}_{2\lfloor v\rfloor}\underbrace{\uparrow\uparrow\cdots\uparrow}_{N-2\lfloor v\rfloor}\rangle\in E_{-v}.
\label{Cv}
\ee
We now outline an algorithm for reaching one of the canonical configurations by repeated applications of $T$ on a general arrow configuration. 

A general arrow configuration can be partitioned into {\em sections} consisting of aligned arrows in such a way that two neighbouring sections contain opposite arrows. 
If the configuration is composed of one section only, it is already in canonical form ((\ref{C0}) for $N$ even; (\ref{Cv}) for $N$ odd), so let us assume it consists of several sections. From the identities
\be
\s_i^- \s_{i+1}^- = \s_i^- \s_{i+1}^- V_1 V_1\,, \qquad \s_i^+ \s_{i+1}^+ = V_1 \s_i^- \s_{i+1}^- V_1\,,
\ee
we see that flipping a pair of adjacent aligned arrows produces another configuration in the same $T$-orbit (in fact $T^2$-orbit), and similarly for an even length sequence of aligned arrows. If the leftmost section in our general configuration contains an even number of arrows, we flip these arrows. The result is a configuration whose (new) leftmost section is enlarged. If this section is even, we repeat the previous flipping step. This is repeated until we have produced a canonical configuration with all arrows aligned or a configuration whose leftmost section contains an {\em odd} number of arrows followed by at least one other section. In the latter case, we may now `move' the interface between the two leftmost sections all the way to the left by application of $\prod_{j=2}^m\s_j^-$ or $\prod_{j=2}^m\s_j^+$, where $m$ is the odd number of arrows in the leftmost section. The result is a configuration in which the two leftmost arrows are opposite. We then repeat the procedure just outlined on the remaining $N-2$ arrows. After a certain number of repetitions, the resulting configuration will be of the form
\be
 |(\downarrow\uparrow)(\downarrow\uparrow)\cdots(\downarrow\uparrow)\uparrow\uparrow\cdots\uparrow\rangle\qquad {\rm or}\qquad
 |(\uparrow\downarrow)(\uparrow\downarrow)\cdots(\uparrow\downarrow)\downarrow\downarrow\cdots\downarrow\rangle.
\label{alg}
\ee
If $N$ is even, the number of non-embraced arrows to the right is even, and flipping these produces  a configuration in canonical form (\ref{Cv}). If $N$ is odd, there is an even number of aligned arrows in the rightmost section, and flipping them produces a canonical configuration of the form (\ref{Cv}). This completes the proof in the open case.

Let us now turn to the periodic case. Because the periodic transfer matrix $T^{\rm per}$ involves an extra factor compared to the open matrix $T^{\rm open}$, namely $T^{\rm per} = [{\mathbb I} + \a\, \s_N^-\s_1^-] \, T^{\rm open}$, it follows that the $T^{\rm per}$-orbits must be unions of the $T^{\rm open}$-orbits, that is, unions of $E_0$ and $E_v \oplus E_{-v}$ for appropriate values of $v>0$.

For $N$ even, the anti-commutation of the transfer matrix with $\cal V$ implies that $T$ cannot connect two sectors with different values of $|v|$. Therefore, the $T$-orbits in the periodic case coincide with those of the open case, and are given by the sectors (\ref{TE}).

For $N$ odd, the previous argument does not apply since the anti-commutation of $T$ and $\cal V$ is lost. It is, however, not difficult to see that the full configuration space forms a single orbit under $T$. Indeed, let us consider the (unique) configuration $|(\uparrow\downarrow)\cdots(\uparrow\downarrow)\!\uparrow\rangle$ in $E_{-n}$. Since its first and last arrows are up, by applying $\s_N^-\s_1^- = \s_N^-\s_1^- V_1V_1$, we obtain a non-zero configuration of $E_{-n+2}$, itself related, by some power of $T^2$, to another configuration of $E_{-n+2}$ with its first and last arrows up (if $-n+2 < n-1$). Iterating these transformations, we see that the subspaces $E_{-n},\, E_{-n+2}, \ldots, E_{n-1}$ form a single orbit under $T^2$. Likewise, using $\s_N^+\s_1^+$ yields another orbit under $T^2$ formed by the subspaces $E_n,\,E_{n-2}, \ldots, E_{-n+1}$. These two $T^2$-orbits are clearly related by a global flip $V_1$ and thus form a single orbit under $T$.

This completes the classification of the $T$-orbits: they coincide with the sectors in (\ref{TE}) in the open cases and in the even periodic case, whereas in the odd periodic case, the full space forms a single $T$-orbit. Consequently, the $T^2$-orbits are given by the $T^2$-sectors in (\ref{T2E}) in the open and even periodic cases, because $T^2$ preserves the value of $v$ in these cases. In the odd periodic case, the above arguments show that the two $T^2$-orbits are given by
\be
\mathcal{E}_+=\bigoplus_{v=-n+1,\,{\rm by}\,2}^n E_v,\qquad \mathcal{E}_-=\bigoplus_{v=-n,\,{\rm by}\,2}^{n-1} E_v.
\label{Eplus}
\ee

In what follows, we determine the conformal content of the various sectors discussed above.

\section{Conformal spectra and partition functions}
\label{Section:ConformalSpectra}

\subsection{Diagonalisation of the transfer matrix}

For the two boundary conditions, periodic or open, it is easy to see that $T$ is symmetric and therefore diagonalisable. For $\alpha\in\mathbb{R}$, $T$ is furthermore real. The explicit diagonalisation may be achieved through a Jordan-Wigner transformation. Because the non-periodic case is slightly simpler, and since the periodic case was treated by Lieb in~\cite{lieb}, we will briefly outline the diagonalisation procedure in the open case. In view of the results of the previous section, it is natural to diagonalise $T^2$ (diagonalising $T$ itself is possible by the same technique, but involves unnecessary complications), 
\be
 T^2 = V^{}_3 V_3^\dagger = \exp{(\a \sum_{i=1}^{N-1}\,\s_i^-\s_{i+1}^-)} \, \exp{(\a \sum_{i=1}^{N-1}\,\s_i^+\s_{i+1}^+)}. 
    \qquad\qquad \hbox{(open case)}
\ee
We start by finding the spectrum of $T^2$. From this, in Section~\ref{Section:SectorsSpectra}, we form the eigenbasis for $\V$. 

Following Lieb, we define the following fermionic operators,
\be
 C^{}_i = (-1)^{i-1} \: (\prod_{k=1}^{i-1} \: \s^z_k) \: \s^-_i\,, \qquad 
 C^\dagger_i = (-1)^{i-1} \: (\prod_{k=1}^{i-1} \: \s^z_k) \: \s^+_i\,, \qquad (1 \leq i \leq N).
\label{jw}
\ee
The algebra satisfied by these operators follows easily from the algebra of the Pauli matrices,
\be
 \{C^{}_i,C^{}_j\} =\{C^\dagger_i,C^\dagger_j\} = 0, \qquad \{C^{}_i,C^\dagger_j\} = \mathbb I \, \delta_{i,j}\,.
\ee

From (\ref{jw}), we readily obtain $C_i \, C_{i+1} = -\s^-_i \, \s^-_{i+1}$, and thus
\be
 T^2 = \exp{(-\a \sum_{i=1}^{N-1}\,C_i C_{i+1})} \; \exp{(\a \sum_{i=1}^{N-1}\,C^\dagger_i C^\dagger_{i+1})}.
\ee
The quadratic forms in the exponentials can be block-diagonalised by the following Fourier transformation,
\be
 C^{}_j = \sqrt{2 \over N+1} \: e^{{\rm i}\gamma} \: \sum_{k=1}^N \; {\rm i}^{j+k} \: \sin{\pi kj \over N+1} \, \eta^{}_k\,, \qquad 
 C^\dagger_j = \sqrt{2 \over N+1} \: e^{-{\rm i}\gamma} \: \sum_{k=1}^N \; {\rm (-i)}^{j+k} \: \sin{\pi kj \over N+1} \, \eta^\dagger_k\,,
 \label{fourier}
\ee
with $e^{2{\rm i}\gamma} = {\rm (-i)}^N$ for convenience. The transformation being unitary, the fermions $\eta_k^{},\eta_k^\dagger$ satisfy the same algebra as the fermionic operators $C^{}_j,C^\dagger_j$. Under this transformation, $T^2$ acquires the factorized form
\be
 T^2 = \bigotimes_{k=1}^{\lfloor{N+1 \over 2}\rfloor}\: A_k, \qquad A_k = \exp{\Big\{2\a\label{pauli}
 \cos{q^{}_k} \; \eta^{}_k \, \eta^{}_{N+1-k}\Big\}} \: \exp{\Big\{2\a\cos{q^{}_k} \; \eta^\dagger_{N+1-k} \, \eta^\dagger_{k}\Big\}},
  \qquad q_k = {\pi k \over N+1}. 
\label{TT}
\ee
We note that for $N$ odd and $k={N+1 \over 2}$, the block $A_k = \bi_2$ is the identity ($\cos q_k = 0$) and acts in the two-dimensional space generated by $|0\ra$ and $\eta^\dagger_{N+1 \over 2}|0\ra$. For  $k \leq {N \over 2}$, each block $A_k$ acts in the four-dimensional Fock space generated by $|0\ra, \, \eta^\dagger_k|0\ra, \, \eta^\dagger_{N+1-k}|0\ra$ and $\eta^\dagger_k\eta^\dagger_{N+1-k}|0\ra$. In all cases, the full space has dimension $2^N$. 

The diagonalisation of the blocks $A_k$ is straightforward and yields the following eigenvalues,
\bea
&& \hspace{-7mm} \lambda_k = 1\,, \; 1\,, \; [\sqrt{1+\a^2 \cos^2{q_k}} + \a \cos{q_k}]^2 \,,\; [\sqrt{1+\a^2 \cos^2{q_k}} - \a \cos{q_k}]^2\,, \qquad (k \leq \textstyle{N \over 2}) \label{eq} \\ 
\noalign{\medskip}
&& \hspace{-7mm} \lambda_k = 1\,,\; 1\,. \qquad\quad (k = \textstyle{N+1 \over 2}\in\mathbb{N})
\eea
In (\ref{eq}), the two eigenvalues $1$ correspond to the odd (fermonic) eigenvectors $\eta^\dagger_k|0\ra$ and $\eta^\dagger_{N+1-k}|0\ra$, while the other two eigenvectors are suitable linear combinations of the even (bosonic) states $|0\ra$ and $\eta^\dagger_k\eta^\dagger_{N+1-k}|0\ra$.

It follows that the eigenvalues of $T^2$ are of the form
\be
 \lambda = \prod_{k=1}^{\lfloor{N \over 2}\rfloor} \, \lambda_k
\ee 
with the possible values of each $\lambda_k$ as listed above. For $N$ odd, each such eigenvalue must be counted twice due to the presence of the block $A_{N+1 \over 2}=\bi_2$. For both parities, the eigenvalues can be conveniently written as
\be 
 \lambda = \prod_{k=1 \atop k=N-1 \bmod 2}^{N-1} \: \Big[\sqrt{1+\a^2 \sin^2{p_k}} + \a \sin{p_k}\Big]^{2(1-\ep_k - \mu_k)}\,, \qquad \quad p_k = {\pi k \over 2(N+1)}\,,
\ee 
where for each $k$, the numbers $\ep_k$ and $\mu_k$ independently take the values 0 or 1. This convention differs from the one used in~\cite{pera} on critical dense polymers.

We finish this section by computing the conformal limit of this spectrum. The energy eigenvalues are  obtained from the above eigenvalues $\lambda$ as 
\be
 E = -{1 \over 2} \log{\lambda}.
\ee
We consider the cases $N$ even and $N$ odd separately.

\subsection{Conformal spectra for \boldmath{$N$} even}

Here we consider the case $N$ even.
Since $\sin{p_k} > 0$ for all $k$, the (non-degenerate) maximal eigenvalue $\lambda_{\rm max}$ is obtained by choosing all $\ep_k = \mu_k = 0$. It follows that the ground-state energy is given by
\bea
E_0 &=& -{1 \over 2} \log{\lambda_{\rm max}} = - \, \sum_{k=1,\,{\rm odd}}^{N-1} \, \log{\Big(\sqrt{1+\a^2 \sin^2{p_k}} + \a \sin{p_k}\Big)} = - \, \sum_{k=1,\,{\rm odd}}^{N-1} {\rm arcsinh} (\alpha \sin{\textstyle {\pi k \over 2(N+1)}})\nonumber\\
&=& -{N+1 \over \pi} \int_0^{\frac{\pi}{2}} {\rm d}t \: {\rm arcsinh}{(\alpha \sin{t})} + \frac{1}{2}\, {\rm arcsinh}\,\alpha - {\alpha \pi \over 24N} + \ldots,
\label{gseven}
\eea
where the last expression indicates the first few terms in the asymptotic expansion of $E_0$ with respect to the system size $N$. Using the convention
\be
 E_0=Nf_{\mathrm{bulk}}+f_{\mathrm{bdy}}+\mathcal{O}({\textstyle \frac{1}{N}})
\ee
to identify the free energies, it follows that the bulk free energy per site (half a dimer) is
\be
 f_{\mathrm{bulk}}= -{1 \over \pi} \int_0^{\frac{\pi}{2}}  {\rm d}t \: {\rm arcsinh}{(\alpha \sin{t})}=-\frac{1}{\pi}\int_0^\alpha {\rm d}t \: \frac{{\rm arctan}(t)}{t}=\frac{\ri}{\pi}\chi^{}_2(\ri\alpha),
\label{fbulk}
\ee
see Appendix~\ref{AppBFE}.
For $\alpha=1$, the Legendre chi function in (\ref{fbulk}) reduces to Catalan's constant,
\be
 -\ri\chi^{}_2(\ri)= G=0.915965594177\ldots .
\ee
Likewise, the (total) boundary free energy is given by
\be
 f_{\mathrm{bdy}}= \frac{1}{2}\, {\rm arcsinh}\,\alpha-{1 \over \pi} \int_0^{\frac{\pi}{2}} \: {\rm d}t \: {\rm arcsinh}{(\alpha \sin{t})} =\frac{1}{2}\, {\rm arcsinh}\,\alpha+\frac{\ri}{\pi}\chi^{}_2(\ri\alpha).
\label{fbdy}
\ee
Interpreting the $\frac{1}{N}$-correction in (\ref{gseven}) as
\be
 - {\alpha \pi \over 24N}=\frac{\alpha\pi}{N}\Big(-\frac{c}{24}+\Delta_0^{\mathrm{even}}\Big),
\ee
we read off the effective central charge
\be
 c_{\mathrm{eff}}=c-24\Delta_0^{\mathrm{even}}=1
\ee
where $\Delta_0^{\mathrm{even}}$ is the conformal weight of the ground state for $N$ even.

The asymptotic expansion for large $N$ of the excited levels reads
\be
 E - E_0 = \sum_{k=1,\,{\rm odd}}^{N-1} \;(\ep_k + \mu_k) \: {\rm arcsinh} (\alpha\sin{\textstyle {\pi k \over 2(N+1)}}) = {\alpha\pi \over 2N} \sum_{k \geq 1,\,{\rm odd}} \; (\ep_k + \mu_k) \: k + \ldots .
\label{speceven}
\ee
The excited energy levels are thus of the form (by redefining $\ep_k,\mu_k \rightarrow \ep_j,\mu_j$ for $k=2j-1$)
\be
 E_\rho = E_0 + {\alpha\pi \over N}\rho,\qquad \rho = \sum_{j \geq 1} \, (\ep_j + \mu_j) \, (j - {1 \over 2}).
\ee 
The generating function for the level degeneracies $d_\rho$ is then given by
\be
 \sum_{\rho=0}^\infty \; d_\rho \, q^\rho = \prod_{j=1}^\infty \: \sum_{\ep_j,\mu_j\in\{0,1\}}  q^{(\ep_j + \mu_j)(j-{1 \over 2})} = \prod_{j=1}^\infty \: (1 + q^{j-{1 \over 2}})^2 = q^{\frac{1}{24}} \: {\theta_3(q) \over \eta(q)},
\ee 
where $\eta(z)$ is the Dedekind eta function (\ref{eta}). Here and in the following, $\theta_j$ is the $j$-th standard Jacobi theta function given in Appendix~\ref{AppTheta}. We obtain the conformal spectrum generating function,
\be
Z(q) = \sum_E \: e^{-EM} = e^{{\alpha\pi M \over 24N}} \sum_{\rho=0}^\infty \:d_\rho \: e^{-{\alpha\pi M \over N}\rho} = {\theta_3(q) \over \eta(q)}, \qquad q=e^{-\frac{\alpha\pi M}{N}}.
\label{zeven}
\ee
Up to a renormalisation,
it is equal to ${\rm Tr}\,(T^2)^{\frac{M}{2}}$ and may be interpreted as the universal partition function of the dimer model on a cylinder of even height $N$ and even perimeter $M$. The result (\ref{zeven}) is well known~\cite{MW73,LuWu99}.

\subsection{Conformal spectra for \boldmath{$N$} odd}

The case $N$ odd is very similar to the even case. The ground-state energy, now doubly degenerate, is 
\bea
E_0 &=& - \sum_{k=2,\,{\rm even}}^{N-1} \; {\rm arcsinh} (\alpha \sin{\textstyle {\pi k \over 2(N+1)}})\nonumber\\
&=& -{N+1 \over \pi} \int_0^{\frac{\pi}{2}} {\rm d}t \: {\rm arcsinh}{(\alpha \sin{t})} + \frac{1}{2}\, {\rm arcsinh}\,\alpha + {\alpha \pi \over 12N} + \ldots .
\label{gsodd}
\eea
The bulk and boundary free energies are as for $N$ even, (\ref{fbulk}) and (\ref{fbdy}), while the ground-state conformal weight for $N$ odd is related to the one for $N$ even by
\be
 \Delta_0^{\mathrm{odd}}=\Delta_0^{\mathrm{even}}+\frac{1}{8}.
\ee

The values of the excited levels are given by
\be
 E - E_0 = \sum_{k=2,\,{\rm even}}^{N-1} (\ep_k + \mu_k) \: {\rm arcsinh} (\alpha\sin{\textstyle {\pi k \over 2(N+1)}}) = {\alpha\pi \over 2N} \sum_{k \geq 2,\, {\rm even}} \; (\ep_k + \mu_k) \: k + \ldots,
\label{specodd}
\ee
each such eigenvalue being counted twice. The excited energy levels are thus of the form (by redefining $\ep_k,\mu_k \rightarrow \ep_j,\mu_j$ for $k=2j$)
\be
 E_\rho = E_0 + {\alpha\pi \over N}\rho,\qquad \rho = \sum_{j \geq 1}\, (\ep_j + \mu_j) \, j. 
\ee
It follows that the generating function for the degeneracies is 
\be
 \sum_{\rho=0}^\infty \; d_\rho \, q^\rho = 2 \prod_{j=1}^\infty \: (1 + 2q^j + q^{2j}) = 2 \prod_{j=1}^\infty \: (1 + q^j)^2 = q^{-\frac{1}{12}} \: {\theta_2(q) \over \eta(q)}.
\ee 
The spectrum generating function, identified with the universal partition function on a cylinder of odd height $N$ and even perimeter $M$, becomes~\cite{MW73,Izm03}
\be
Z(q) = \sum_E \: e^{-EM} = e^{-{\alpha\pi M \over 12N}} \sum_{\rho=0}^\infty \:d_\rho \: e^{-{\alpha\pi M \over N}\rho} = {\theta_2(q) \over \eta(q)}, \qquad q=e^{-\frac{\alpha\pi M}{N}}.
\label{zodd}
\ee

\section{Sectors and refined spectra}
\label{Section:SectorsSpectra}

Using the definition (\ref{jw}) of the auxiliary fermions $C^{}_j,C^\dagger_j$ in terms of Pauli matrices, and then the Fourier transform (\ref{fourier}) to the fermions $\eta_k,\eta_k^\dagger$, the operator $\V$ can be written as\be
 \V = {1 - (-1)^N \over 4} \, + \: \sum_{j=1}^N \,(-1)^j \, C^\dagger_j C^{}_j
= {1 - (-1)^N \over 4} \, + \: {\rm i}^{N-1} \: \sum_{k=1}^N \: (-1)^k \, \eta^\dagger_k \, \eta^{}_{N+1-k}.
\label{Gsum}
\ee
Now, recalling that $T^2$ factorizes as
\be
T^2 = \bigotimes_{k=1}^{\lfloor{N+1 \over 2}\rfloor}\: A_k, \qquad A_k = \exp{\Big\{2\a \, \cos{q_k} \; \eta^{}_k \, \eta^{}_{N+1-k}\Big\}} \: \exp{\Big\{2\a \, \cos{q_k} \; \eta^\dagger_{N+1-k} \, \eta^\dagger_{k}\Big\}},
\label{TT2}
\ee
we see that the terms in the sum (\ref{Gsum}) should be grouped in pairs to act on the same spaces as the blocks $A_k$. This allows us to examine each block separately. As seen in the following, the action depends slightly on the parity of $N$. Before examining the spectrum in the various sectors, we discuss the relation between the variation index and the fermionic parity.

\subsection{Variation index and the fermion number}

The Jordan-Wigner transformation converts the arrow configuration space into a fermionic Fock space whose states are built by applying the discrete fermions $\eta^\dagger_k$ on the vacuum state $|0\ra$. The vacuum, being annihilated by all $\eta^{}_k$, hence by all $C_i$, is proportional to $|\!\!\downarrow \downarrow \cdots \downarrow \rangle$. The full space may be split into an even sector and an odd sector, made up respectively of states containing an even or odd number of fermionic excitations. From the explicit expression (\ref{TV3V1}) of the transfer matrix, and the relations (\ref{jw}) between the Pauli matrices and the fermions, we see that $T$ leaves the even and odd sectors invariant if $N$ is even, but interchanges them if $N$ is odd. The same is true in the periodic case \cite{lieb}. It follows that the even and odd sectors are left invariant by $T^2$. The above expression (\ref{TT2}) makes this manifest in the open case.  

In all cases, open or periodic, the fermion number operator, given by
\be
{\cal N} = \sum_{k=1}^N \: \eta^\dagger_k\,\eta^{}_k = \sum_{i=1}^N\: C^\dagger_i \, C_i = \sum_{i=1}^N \: \sigma^+_i \sigma^-_i,
\ee
is closely related to the variation index operator $\V$. In particular, the fermionic parity $(-1)^{\cal N}$ is related to the variation index by
\be
(-1)^{\cal N} = \cases{
(-1)^\V & if $N$ is even,\cr
(-1)^{\V - {1 \over 2}} & if $N$ is odd.}
\ee

In the periodic case with $N$ odd, the two invariant sectors $\mathcal{E}_+$ and $\mathcal{E}_-$ coincide with the even and odd subspaces, although the precise identification depends on $N$. Since the vacuum state, which is an even state, has variation index $v=+\half$, we obtain from (\ref{Eplus}) that the even subspace is equal to $\mathcal{E}_+$ for $N=1 \bmod 4$, and to $\mathcal{E}_-$ for $N=3 \bmod 4$.

\subsection{Refined spectra for \boldmath{$N$} even}

For $N$ even, the operator $\V$ is rewriten as
\be
 \V = {\rm i}^{N-1} \: \sum_{k=1}^{\frac{N}{2}} \: (-1)^k \, \Big[\eta^\dagger_k \, \eta^{}_{N+1-k} - \eta^\dagger_{N+1-k} \, \eta^{}_{k}\Big]   \equiv \sum_{k=1}^{\frac{N}{2}} \: \V_k,
\ee
thereby introducing the operators $\V_k$ as the indicated summands. For every $k$, the operator $\V_k$ is diagonalisable and commutes with the corresponding block $A_k$. It follows that the two non-degenerate eigenvectors of $A_k$ are eigenvectors of $\V_k$ as well. Indeed, we find 
\be
 \V_k\,|0\ra = \V_k \,\eta^\dagger_k\eta^\dagger_{N+1-k}\,|0\ra = 0.
\ee
Also, it is recalled that the two eigenvectors of $A_k$ corresponding to the degenerate eigenvalue 1 are $\eta^\dagger_k\,|0\ra$ and $\eta^\dagger_{N+1-k}\,|0\ra$. In the basis of these two vectors, $\V_k$ is represented by the matrix 
\be
 \V_k = {\rm i}^{N-1} \, (-1)^k \, \pmatrix{0 & 1 \cr -1 & 0},
\ee
with eigenvalues $v_k = \pm \ci^N (-1)^k\in\{-1,1\}$. The degenerate $A_k$ eigenspace of eigenvalue $1$ is therefore spanned by two $\V_k$ eigenvectors with the distinct eigenvalues $+1$ and $-1$.

Returning to the spectrum of $A_k$ given earlier, namely
\be
\lambda_k = \Big[\sqrt{1+\a^2 \sin^2{p_k}} + \a \sin{p_k}\Big]^{2(1-\ep_k - \mu_k)}\,, 
\ee
we see that the two states with $\ep_k + \mu_k = 0,2$ have $v_k=0$ whereas those with $\ep_k + \mu_k = 1$ have $v_k = \pm 1$. Introducing an extra parameter $y$ to keep track of the eigenvalues of $\V$, we obtain the following generating function for the degeneracies $d_{\rho,v}$ at level $\rho$ and eigenvalue $v$ of $\V$,
\be
\sum_{\rho=0}^\infty \: \sum_{v \in \Z} \: d_{\rho,v} \, q^\rho \, y^v = \prod_{j=1}^\infty \: \Big(1 + (y + y^{-1}) \, q^{j-{1 \over 2}} + q^{2j-1}\Big) = q^{\frac{1}{24}} \: {\theta_3(y|q) \over \eta(q)}.
\label{genfunceven}
\ee 
Taking the ground-state energy into account, the corresponding conformal partition function reads
\be
Z(q;y) = {\rm Tr}\,(T^M y^\V) = q^{-\frac{1}{24}}\: \sum_{\rho=0}^\infty \: \sum_{v \in \Z} \:d_{\rho,v} \: q^{\rho} \, y^v = {\theta_3(y|q) \over \eta(q)}, \qquad q=e^{-\frac{\alpha\pi M}{N}}.
\ee
By expanding $\theta_3$ in a power series in $y$ and comparing with 
\be
 Z(q;y) = \sum_v \, Z_v(q) \, y^v,
\ee 
we obtain the partition functions pertaining to the individual $T^2$-sectors,
\be
Z_v(q) = {q^{\frac{v^2}{2}} \over \eta(q)}\,, \qquad\qquad v \in \mathbb Z.
\label{Zveven}
\ee
It is recalled that this derivation of the sector partition functions (\ref{Zveven}) is based on $T^2$, so $M$ is assumed even.

\subsection{Refined spectra for \boldmath{$N$} odd}

For $N$ odd, the operator $\V$ in (\ref{Gsum}) may be written as
\be
\V = \Big[{1 \over 2} - \eta^\dagger_{\frac{N+1}{2}} \, \eta^{}_{\frac{N+1}{2}}\Big] + \ci^{N-1} \: \sum_{k=1}^{\frac{N-1}{2}} \: (-1)^k \, \Big[\eta^\dagger_k \, \eta^{}_{N+1-k} + \eta^\dagger_{N+1-k} \, \eta^{}_{k}\Big] \nonumber\\
\equiv \V_{\frac{N+1}{2}} + \sum_{k=1}^{\frac{N-1}{2}} \: \V_k,
\ee
thereby introducing the (diagonalisable) operators $\V_k$. For every $k\leq\frac{N+1}{2}$, the operator $\V_k$ acts in the same space as the block $A_k$ with which it commutes. Both $A_{\frac{N+1}{2}} = \bi_2$ and $\V_{\frac{N+1}{2}}$ are two-dimensional.

For $k \leq \frac{N-1}{2}$, as before, we find that the non-degenerate eigenvectors of $A_k$ have $v_k = 0$. Moreover, on the two degenerate states $\eta^\dagger_k\,|0\ra$ and $\eta^\dagger_{N+1-k}\,|0\ra$, $\V_k$ is represented by
\be
\ci^{N-1} \, (-1)^k \, \pmatrix{0 & 1 \cr 1 & 0},
\ee
with eigenvalues $v_k = \pm \ci^{N-1}(-1)^k\in\{-1,1\}$. We reach the same conclusion as in the $N$ even case: the two eigenvectors of $A_k$ with eigenvalue 1 have $v_k = \pm 1$. 

In the remaining two-dimensional block, the operator $\V_{\frac{N+1}{2}} = {1 \over 2} - \eta^\dagger_{\frac{N+1}{2}} \, \eta^{}_{\frac{N+1}{2}}$ takes respectively the values $1 \over 2$ and $-{1 \over 2}$ on the states $|0\ra$ and $\eta^\dagger_{\frac{N+1}{2}}|0\ra$, themselves eigenvectors of $A_{\frac{N+1}{2}}$ with eigenvalue 1. While in the previous section, this doubly degenerate eigenvalue 1 of $A_{\frac{N+1}{2}}$ caused the duplication of all eigenvalues of $T^2$, it now gives an extra contribution $v_{\frac{N+1}{2}} = \half$ to half the spectrum, and $v_{\frac{N+1}{2}} = -\half$ to the other half.

The corresponding generating function for the degeneracies $d_{\rho,v}$ reads
\be
\sum_{\rho=0}^\infty \: \sum_{v \in \Z+\frac{1}{2}} \: d_{\rho,v} \, q^\rho \, y^v = (y^{\frac{1}{2}} + y^{-\frac{1}{2}}) \: \prod_{j=1}^\infty \: \Big(1 + (y + y^{-1}) \, q^j + q^{2j}\Big) = q^{-\frac{1}{12}} \: {\theta_2(y|q) \over \eta(q)},
\label{genfuncodd}
\ee 
and leads to the following conformal partition function,
\be
Z(q;y) = {\rm Tr}\,(T^M y^\V) = q^{\frac{1}{12}}\: \sum_{\rho=0}^\infty \: \sum_{v \in \Z+\frac{1}{2}} \:d_{\rho,v} \: q^{\rho} \, y^v = {\theta_2(y|q) \over \eta(q)}, \qquad q=e^{-\frac{\alpha\pi M}{N}}.
\ee
From the power expansion of $\theta_2$ with respect to $y$, we obtain the partition functions pertaining to the individual $T^2$-sectors,
\be
Z_v(q) = {q^{\frac{v^2}{2}} \over \eta(q)}\,, \qquad\qquad v \in \mathbb Z + \half,
\ee
again assuming $M$ even.

Strikingly, for every $N$, even or odd, we obtain the unified conclusion that in the conformal limit, the sectors are labelled by (half-)integers $v\in\Z + {N \over 2}$, and the partition function of a single sector is equal to the character of a single Verma module,
\be
Z_v(q) = Z_{-v}(q) = {q^{{v^2 \over 2}-{1 \over 24}} \over \displaystyle{\prod_{m=1}^\infty (1-q^m)}}.
\label{ZvZv}
\ee
However, as discussed in the following, we do {\em not} claim that the underlying representations necessarily correspond to Verma modules, only that the characters are as given in (\ref{ZvZv}).

\section{Conformal field theory}
\label{SecCFT}

It is natural to ask what conformal representations arise in the continuum scaling limit of the various sectors. As in other comparable studies, the answer is not immediate from the knowledge of the transfer matrix alone. In particular, the possibility of non-trivial Jordan blocks in the conformal limit is a subtle issue. The transfer matrix under investigation here is diagonalisable. However, many eigenvalues, which are distinct for finite system sizes, become degenerate in the conformal spectrum, thereby making room for non-trivial Jordan blocks. Thus, the diagonalisability of a finite-size transfer matrix does not necessarily preclude the existence of such Jordan blocks in the conformal representations, although no example of this kind seems to be known\footnote{In the critical dense polymer model, for instance, the non-trivial Jordan blocks are present for finite system sizes~\cite{pera,mdsa,vjs}.}$\!$. We intend to address this possibility for the dimer model elsewhere. In addition, as is well-known, the value of the central charge $c$ is crucial to characterise the associated conformal field theory and the structure of the various representations, but this value cannot be determined unequivocally from the spectrum alone. 

Here we will make the ansatz that the dimer model has 
\be
 c=-2.
\label{c-2}
\ee 
This value is borne out by a number of independent observations and results: {(i)} the dependence of the effective central charge on the parity of the width of a strip is easily and naturally explained from that particular value of $c$~\cite{iprh}; {(ii)} there is a well-known equivalence between the dimer model in a rectangular geometry, the uniform spanning tree model and the Abelian sandpile model, the latter providing overwhelming evidence that $c=-2$~\cite{maru,ru1,jpr,ru2,ppr}; {(iii)} the partition function on a rectangle with odd side lengths and one corner site removed is equal to $\eta(q)$ in agreement with the general result $[\eta(q)]^{-c/2}$, see~\cite{kleva}; and {(iv)} the finite-size corrections to the conformal spectrum can be reproduced in perturbation theory for $c=-2$, but appear unreproducible for $c=1$ (as we will discuss elsewhere). Moreover, we will see that the conformal spectrum of the dimer model is very closely related to that in the critical dense polymer model, also believed to have $c=-2$, see~\cite{pera,PRV}, for example.

Notwithstanding, based on different observations, there are also strong arguments favouring the other widely proposed value, namely $c=1$: {(i)} the so-called height function, defined on dimer configurations, converges in the scaling limit to a Gaussian free field~\cite{ken}; and {(ii)} in the generalised model allowing for monomers in addition to dimers, it has been shown that the monomer correlators are those of free complex fermions, both on a boundary~\cite{pr} and in the bulk~\cite{dub}, following earlier observations~\cite{aupe,fms,plf}.

Taken together, these results and observations seem to indicate that the pure dimer model has $c=-2$, while the monomer-dimer model has $c=1$. In this scenario, the height function should be viewed as belonging to the monomer-dimer model.

Now, an immediate consequence of assuming $c=-2$ is that the parity-dependent ground-state conformal weights are
\be
 \Delta_0^{{\rm even}}=-{\textstyle \frac{1}{8}},\qquad \Delta_0^{{\rm odd}}=0,
\ee
while the representation associated with the partition function $Z_v(q)$ in the sector $v$ has conformal weight 
\be
 \Delta(v) = {4v^2 - 1 \over 8},\qquad v\in\half\mathbb{Z}.
\label{Dv}
\ee 
These weights are integer if $N$ is odd (for which $v\in\mathbb{Z}+\half$), but equal to $-{1 \over 8}$ modulo an integer or a half-integer if $N$ is even (for which $v\in\mathbb{Z}$). They can be expressed by the Kac formula for $c=-2$ as
\be
 \Delta_{r,s}=\frac{(2r-s)^2-1}{8},\qquad r,s\in\mathbb{N},
\ee
where $r,s$ are Kac labels. The corresponding extended Kac table is shown in Figure~\ref{Kac}. Specifically, the set of conformal weights in (\ref{Dv}) are covered exactly once by setting
\be
 |v|=r-\frac{s}{2},\qquad r\in\mathbb{N},\quad s\in\{1,2\}.
\ee
This corresponds to the two lower rows in the Kac table. Following~\cite{Sal92,PRV}, and without claiming any superconformal symmetry in the dimer model, we refer to them as the Neveu-Schwarz ($s=1$) and Ramond ($s=2$) sectors. Alternatively, the conformal weights in (\ref{Dv}) are covered exactly once by setting
\be
 |v|=\frac{s}{2}-r,\qquad r=1,\quad s=2,3,\ldots,
\label{valt}
\ee
noting that $\Delta_{1,1}=\Delta_{1,3}=0$.
As discussed below, this labelling offers a direct link to the characterisation of a particular class of boundary conditions in the critical dense polymer model~\cite{pera}.

\begin{figure}
{\vspace{0in}\psset{unit=.7cm}
{
\small
\begin{center}
\qquad
\begin{pspicture}(0,0)(7,11)
\psframe[linewidth=0pt,fillstyle=solid,fillcolor=lightblue](0,0)(7,11)
\multiput(0,-1)(0,1){2}{\psframe[linewidth=0pt,fillstyle=solid,fillcolor=midblue](0,1)(7,2)}
\psframe[linewidth=0pt,fillstyle=solid,fillcolor=midlightblue](0,2)(1,11)
\psgrid[gridlabels=0pt,subgriddiv=1]
\rput(.5,10.65){$\vdots$}\rput(1.5,10.65){$\vdots$}\rput(2.5,10.65){$\vdots$}
\rput(3.5,10.65){$\vdots$}\rput(4.5,10.65){$\vdots$}\rput(5.5,10.65){$\vdots$}
\rput(6.5,10.5){$\vvdots$}\rput(.5,9.5){$\frac{63}8$}\rput(1.5,9.5){$\frac{35}8$}
\rput(2.5,9.5){$\frac{15}8$}\rput(3.5,9.5){$\frac{3}8$}\rput(4.5,9.5){$-\frac 18$}
\rput(5.5,9.5){$\frac{3}8$}\rput(6.5,9.5){$\cdots$}
\rput(.5,8.5){$6$}\rput(1.5,8.5){$3$}\rput(2.5,8.5){$1$}\rput(3.5,8.5){$0$}
\rput(4.5,8.5){$0$}\rput(5.5,8.5){$1$}\rput(6.5,8.5){$\cdots$}
\rput(.5,7.5){$\frac{35}8$}\rput(1.5,7.5){$\frac {15}8$}\rput(2.5,7.5){$\frac 38$}
\rput(3.5,7.5){$-\frac{1}8$}\rput(4.5,7.5){$\frac 38$}\rput(5.5,7.5){$\frac{15}8$}
\rput(6.5,7.5){$\cdots$}\rput(.5,6.5){$3$}\rput(1.5,6.5){$1$}\rput(2.5,6.5){$0$}\rput(3.5,6.5){$0$}
\rput(4.5,6.5){$1$}\rput(5.5,6.5){$3$}\rput(6.5,6.5){$\cdots$}
\rput(.5,5.5){$\frac{15}8$}\rput(1.5,5.5){$\frac {3}{8}$}\rput(2.5,5.5){$-\frac 18$}
\rput(3.5,5.5){$\frac{3}{8}$}\rput(4.5,5.5){$\frac {15}8$}\rput(5.5,5.5){$\frac{35}{8}$}
\rput(6.5,5.5){$\cdots$}\rput(.5,4.5){$1$}\rput(1.5,4.5){$0$}\rput(2.5,4.5){$0$}
\rput(3.5,4.5){$1$}\rput(4.5,4.5){$3$}\rput(5.5,4.5){$6$}\rput(6.5,4.5){$\cdots$}
\rput(.5,3.5){$\frac 38$}\rput(1.5,3.5){$-\frac 18$}\rput(2.5,3.5){$\frac 38$}
\rput(3.5,3.5){$\frac{15}8$}\rput(4.5,3.5){$\frac{35}8$}\rput(5.5,3.5){$\frac{63}8$}
\rput(6.5,3.5){$\cdots$}\rput(.5,2.5){$0$}\rput(1.5,2.5){$0$}\rput(2.5,2.5){$1$}\rput(3.5,2.5){$3$}
\rput(4.5,2.5){$6$}\rput(5.5,2.5){$10$}\rput(6.5,2.5){$\cdots$}
\rput(.5,1.5){$-\frac 18$}\rput(1.5,1.5){$\frac 38$}\rput(2.5,1.5){$\frac{15}8$}
\rput(3.5,1.5){$\frac{35}8$}\rput(4.5,1.5){$\frac{63}8$}\rput(5.5,1.5){$\frac{99}8$}
\rput(6.5,1.5){$\cdots$}\rput(.5,.5){$0$}\rput(1.5,.5){$1$}\rput(2.5,.5){$3$}\rput(3.5,.5){$6$}
\rput(4.5,.5){$10$}\rput(5.5,.5){$15$}\rput(6.5,.5){$\cdots$}
{\color{blue}
\rput(.5,-.5){$1$}
\rput(1.5,-.5){$2$}
\rput(2.5,-.5){$3$}
\rput(3.5,-.5){$4$}
\rput(4.5,-.5){$5$}
\rput(5.5,-.5){$6$}
\rput(6.5,-.5){$r$}
\rput(-.5,.5){$1$}
\rput(-.5,1.5){$2$}
\rput(-.5,2.5){$3$}
\rput(-.5,3.5){$4$}
\rput(-.5,4.5){$5$}
\rput(-.5,5.5){$6$}
\rput(-.5,6.5){$7$}
\rput(-.5,7.5){$8$}
\rput(-.5,8.5){$9$}
\rput(-.5,9.5){$10$}
\rput(-.5,10.5){$s$}}
\end{pspicture}
\end{center}}}
\caption{\label{Kac} Extended Kac table for $c=-2$. The two lower rows label the Neveu-Schwarz $(s=1)$ and Ramond $(s=2)$ sectors.}
\end{figure}

With the ansatz (\ref{c-2}), none of the representations associated with the $v$-sectors is irreducible. Indeed, in the case of open boundary conditions, we have the decompositions
\bea
Z_v(q) &=& {q^{\frac{v^2}{2}} \over \eta(q)} = \sum_{j=0}^\infty \; \ch_{{4(|v|+2j)^2-1 \over 8}}(q)= \sum_{j=0}^\infty \; \ch_{\Delta(|v|+2j)}^{}(q), \qquad\quad v \in \mathbb Z,\qquad \qquad ({\rm Ramond})
\label{vermaeven}\\
Z_v(q) &=& {q^{\frac{v^2}{2}} \over \eta(q)} = \sum_{j=0}^\infty \; \ch_{{4(|v|+j)^2-1 \over 8}}(q)= \sum_{j=0}^\infty \; \ch_{\Delta(|v|+j)}^{}(q), \qquad v \in \mathbb Z +\half,\qquad ({\rm Neveu-Schwarz})\ \
\label{vermaodd}
\eea
in terms of irreducible $c=-2$ characters,
\bea
 \ch_{\Delta(m)}^{}(q)&=&\frac{q^{\frac{m^2}{2}}(1-q^{2|m|+2})}{\eta(q)},\qquad m\in\mathbb{Z},\\
 \ch_{\Delta(m)}^{}(q)&=&\frac{q^{\frac{m^2}{2}}(1-q^{|m|+\frac{1}{2}})}{\eta(q)},\qquad m\in\mathbb{Z}+\half.
\eea
Here we have adopted the notation $\ch_\Delta^{}(q)$ used in~\cite{pera} to indicate irreducible $c=-2$ characters. We will also use the alternative notation introduced in~\cite{pera}, 
\be
 \ch^{}_{r,s}(q)=\ch^{}_{\Delta_{r,s}}(q),
\ee
utilising the Kac labels $r,s$.

The states in the representations underlying the partition functions in (\ref{vermaeven}) and (\ref{vermaodd}) correspond to eigenvectors of the transfer matrix. The combinatorics encapsulating this correspondence is closely related to the physical combinatorics of the critical dense polymer model as described in~\cite{pera,PRV}. This intriguing observation is outlined below and depends on the parity of $N$.

Here we note that the partition functions in (\ref{vermaeven}) and (\ref{vermaodd}) are characters of Verma modules of highest weight $\Delta(v)$. We stress that this does not necessarily imply that the underlying modules are Verma modules, only that they have Verma module characters. In fact, it is beyond the scope of the present work to characterise the specific module structures.

\section{Open boundary conditions}
\label{SecOpen}

\subsection{Physical combinatorics in the Ramond sector: \boldmath{$N$} even}
\label{SecRamond}

From Section~\ref{Section:ConformalSpectra}, with $c=-2$ and $\Delta_0^{{\rm even}}=-\frac{1}{8}$, the conformal energy levels for $N$ even read
\be
 E = E_0 + {\alpha\pi \over N} \sum_{j \geq 1} \; (j-\half) \:\delta_j = {\alpha\pi \over N}\Big\{-{c \over 24} - {1 \over 8} + \sum_{j \geq 1} \; (j-\half)\,\delta_j \Big\}\,,
\label{eneven}
\ee
where $\delta_j = \ep_j + \mu_j = 0,1,1$ or 2. The results of Section~\ref{Section:SectorsSpectra} show that $\delta_j=0,2$ do not contribute to the value of the variation index $v$, while $\delta_j=1$ contributes $+1$ or $-1$ to $v$. 
Our convention will be that $\ep_j=1$ contributes $-1$ to $v$, while $\mu_j=1$ contributes $+1$, so that the value of $v$ of a state specified by the set $\{\ep_j,\mu_j\}_{j \geq 1}$ is given by 
\be
 v =\sum_{j\geq1} \, (\mu_j-\ep_j).
\label{vme}
\ee 
For finite excitations, the upper bound in this summation is given by $n=\frac{N}{2}$.

From (\ref{vermaeven}), the partition functions of the first few $v$-sectors read
\bea
&& Z_0(q) = \mbox{\boldmath${\ch_{-\frac{1}{8}}}(q)$} + \ch_{\frac{15}{8}}(q) + \ch_{\frac{63}{8}}(q) +\ch_{\frac{143}{8}}(q) + \ldots, \label{z0} \\
\noalign{\medskip}
&& Z_1(q) = \mbox{\boldmath${\ch_{\frac{3}{8}}}(q)$} + \ch_{\frac{35}{8}}(q) + \ch_{\frac{99}{8}}(q) +\ch_{\frac{195}{8}}(q) + \ldots, \qquad\quad Z_{-1}(q) = Z_1(q), \label{z1} \\
\noalign{\medskip}
&& Z_2(q) = \mbox{\boldmath${\ch_{\frac{15}{8}}}(q)$} + \ch_{\frac{63}{8}}(q) + \ch_{\frac{143}{8}}(q) +\ch_{\frac{255}{8}}(q) + \ldots, \qquad\, Z_{-2}(q) = Z_2(q).
\label{z2}  
\eea
The reason for indicating some of the characters in boldface is described in the following.

There is a striking similarity between the spectrum (\ref{eneven}) and the spectrum of the critical dense polymer (CDP) model studied in~\cite{pera,PRV}. In fact, the conformal energies (\ref{eneven}) are exactly those of the CDP model on a strip of {\em odd} width (rather than {\em even} as here) provided the values $\{\ep_j,\mu_j\}_{j \geq 1}$ obey certain selection rules. In other words, the CDP and dimer conformal spectra differ only by the {\em degeneracies} of the energy levels. This observation also applies to the Neveu-Schwarz sectors where $v\in\mathbb{Z}+\half$. In addition, a significant part of the CDP conformal spectrum can also be partitioned into sectors\footnote{Although we use the word ``sector'' in the context of the CDP model, as in~\cite{pera}, it does not have the same meaning as in the dimer model. As discussed above, the variation index $v$ is a good quantum number in the dimer model. In the CDP model, on the other hand, the defect number $\ell$ is {\em not} a good quantum number since the corresponding subspaces are {\em not} left invariant by the finite-size transfer matrices. Instead, the CDP transfer matrices are upper block triangular with respect to $\ell$~\cite{pera}.}$\!$, labelled by a non-negative integer $\ell$ counting the number of so-called {\em defects}. This defect number is related to the dimer sectors with $v\geq-\half$ through
\be 
 \ell=2v+1, \qquad v\in\half\mathbb{N}_0-\half.
\label{l2v}
\ee 
This relation is intimately associated with (\ref{valt}). The equations (\ref{z0})-(\ref{z2}) show the conformal content of the first few Ramond sectors in the dimer model; the characters in boldface are those which appear in the CDP model for the corresponding defect number. For instance, the irreducible characters of conformal weights $-\frac{1}{8}, \, \frac{3}{8},\, \frac{15}{8}$ appear respectively once, twice and three times in the dimer model, but only once in the CDP model as $\ell$ varies over the odd positive integers.

The question still remains of how to associate the eigenvalues in (\ref{eneven}) to the various representations. This may be answered in a way reminiscent of the corresponding results in the CDP model, but in a somewhat surprising manner as it is based on the CDP model defined not only on the {\em strip}~\cite{pera}, but also on the {\em cylinder}~\cite{PRV}.

\begin{figure}
\psset{unit=.6cm}
\setlength{\unitlength}{.6cm}
\begin{center}
\begin{pspicture}(0,0.2)(2,7.5)
\psframe[linewidth=0.0pt,linecolor=yellow!40!white,fillstyle=solid,fillcolor=yellow!40!white](0,0)(2,7)
\psline[linewidth=0.5pt](0,7.5)(0,0)(2,0)(2,7.5)
\psarc[linecolor=black,linewidth=.5pt,fillstyle=solid,fillcolor=white](0.5,6.5){.1}{0}{360}
\psarc[linecolor=gray,linewidth=0pt,fillstyle=solid,fillcolor=gray](0.5,5.5){.1}{0}{360}
\psarc[linecolor=black,linewidth=.5pt,fillstyle=solid,fillcolor=white](0.5,4.5){.1}{0}{360}
\psarc[linecolor=black,linewidth=.5pt,fillstyle=solid,fillcolor=white](0.5,3.5){.1}{0}{360}
\psarc[linecolor=gray,linewidth=0pt,fillstyle=solid,fillcolor=gray](0.5,2.5){.1}{0}{360}
\psarc[linecolor=gray,linewidth=0pt,fillstyle=solid,fillcolor=gray](0.5,1.5){.1}{0}{360}
\psarc[linecolor=black,linewidth=0.5pt,fillstyle=solid,fillcolor=white](0.5,0.5){.1}{0}{360}
\psarc[linecolor=black,linewidth=.5pt,fillstyle=solid,fillcolor=white](1.5,6.5){.1}{0}{360}
\psarc[linecolor=black,linewidth=.5pt,fillstyle=solid,fillcolor=white](1.5,5.5){.1}{0}{360}
\psarc[linecolor=gray,linewidth=0pt,fillstyle=solid,fillcolor=gray](1.5,4.5){.1}{0}{360}
\psarc[linecolor=gray,linewidth=0pt,fillstyle=solid,fillcolor=gray](1.5,3.5){.1}{0}{360}
\psarc[linecolor=black,linewidth=.5pt,fillstyle=solid,fillcolor=white](1.5,2.5){.1}{0}{360}
\psarc[linecolor=gray,linewidth=0pt,fillstyle=solid,fillcolor=gray](1.5,1.5){.1}{0}{360}
\psarc[linecolor=gray,linewidth=0pt,fillstyle=solid,fillcolor=gray](1.5,0.5){.1}{0}{360}
\rput(-1,0.45){{\scriptsize $j=1$}}
\rput(-1,1.45){{\scriptsize $j=2$}}
\rput(-1,2.45){{\scriptsize $j=3$}}
\rput(-1,3.45){{\scriptsize $j=4$}}
\rput(-1,4.45){{\scriptsize $j=5$}}
\rput(-1,5.45){{\scriptsize $j=6$}}
\rput(-1,6.45){{\scriptsize $j=7$}}
\rput(0.5,-0.6){$\varepsilon$}
\rput(1.5,-0.65){$\mu$}
\end{pspicture}
\end{center}
\caption{\label{exa}Two-column configuration characterised by non-vanishing occupation numbers $\ep_2,\ep_3,\ep_6,\mu_1,\mu_2,\mu_4,\mu_5$. This is an example of a non-admissible configuration.}
\end{figure}
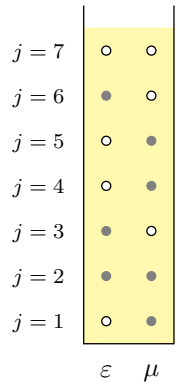

Following~\cite{pera}, we thus associate an eigenvalue of the form (\ref{eneven}) with a two-column (double-column) semi-infinite configuration as in Figure~\ref{exa}, where the levels, labelled by $j \geq 1$ from below, contain two sites, empty (white) or occupied (grey). For a given set $\{\ep_j,\mu_j\}_{j \geq 1}$, the left site at level $j$ is occupied if $\ep_j=1$ but empty if $\ep_j=0$, while the right site at level $j$ is occupied if $\mu_j=1$ but empty if $\mu_j=0$. Finite excitations of the form (\ref{eneven}) correspond to configurations with only finitely many occupants. As indicated, we occasionally refer to an occupied site as an occupant.

For finite system size $N=2n$ and variation index $v$, a natural finitization of the partition function $Z_v(q)$ is now given by
\be
 Z_v^{(N)}(q)=q^{-\frac{c}{24}+\Delta_0^{\mathrm{even}}}\sum_{{\mbox{{\scriptsize $v$--two}} \atop \mbox{{\scriptsize columns}}}}q^{\sum_j \delta_jE_j},
\label{ZvNR}
\ee
where the sum is over all two-column configurations of maximal height $n$ and variation index $v$, and where the elementary excitation energy of an occupant at level $j$ follows from (\ref{eneven}) and is given by
\be
 E_j=j-\half.
\label{EjR}
\ee

The two-column configuration in Figure~\ref{exa}, for example, has energy excitation $E-E_0 = {39 \over 2}\cdot{\alpha\pi \over N}$ and belongs to the sector $v=1$. The non-zero values of $\delta_j = \ep_j+\mu_j$ are $\delta_1=\delta_3=\delta_4=\delta_5=\delta_6=1$ and $\delta_2=2$ implying that this state is degenerate with 31 other states with the same energy and values of $\delta_j$. The corresponding two-column configurations are obtained from the present one by moving occupants from left to right, or from right to left, at the levels containing a single occupant. These 31 other states do not all belong to the same sector, though, since moving an occupant horizontally changes the value of $v$. Each of these states will have a variation index $v$ between $-5$ and $+5$, since there are $5$ distinct levels with a single occupied site. This also accounts for the total of $2^5=32$ related configurations above.

According to~\cite{pera}, a two-column configuration with a finite number of occupants (corresponding to a finite energy excitation) is admissible or non-admissible\footnote{We have chosen to adopt this terminology from the work~\cite{pera} on the CDP model where it is very appropriate. In the dimer model, on the other hand, non-admissible configurations actually carry physical information about states in sub-leading characters. A more befitting name for these configurations could therefore be {\em subordinate} configurations.}$\!$. An {\em admissible} configuration is such that if one draws a horizontal line at any height, the number of occupants on or above the line in the left column must be smaller than or equal to the number of occupants on or above the line in the right column. Equivalently, admissible configurations have values $\{\ep_j,\mu_j\}$ such that 
\be
 \sum_{j \geq\kappa} \:(\mu_j - \ep_j) \geq 0,\qquad \forall \kappa \in\mathbb{N}.
\label{adm}
\ee 
In particular, they all have a smaller or equal number of occupants in the left column than in the right one, and consequently have values $v \geq 0$. It is noted, though, that a two-column with $v\geq0$ need not be admissible, as illustrated in Figure~\ref{exa}. 

The set of admissible configurations of maximal height $h$ (i.e.$\!\,$ no occupants above level $h$) and given variation index $v$ may be decomposed into a disjoint union of subsets, 
\be
 \mathcal{A}_v^{(h)}=\bigcup_{m=0}^{h-v} A^{(h)}_{m,m+v},
\ee 
where $A^{(h)}_{m,m+v}$ contains the admissible configurations of maximal height $h$ with exactly $m$ occupants in the left column and exactly $m+v$ occupants in the right column. Thus, $\mathcal{A}_v^{(h)}$ is the set of admissible configurations which have an excess of $v$ occupants in the right column over the left column. In the continuum scaling limit, where there is no upper bound on $h$, the set of such admissible configurations becomes
\be
 \mathcal{A}_v = \lim_{h\to\infty}\mathcal{A}_v^{(h)}.
\ee

For the CDP model, Pearce and Rasmussen~\cite{pera} conjectured a set of selection rules positing that the CDP {\em conformal} spectrum on a strip of {\em odd} width is given by (\ref{eneven}) for all those values $\{\ep_j,\mu_j\}$ which define admissible two-column configurations. This conjecture has been recently proven by Morin-Duchesne~\cite{morin}. In the continuum scaling limit, the combinatorics of admissible configurations is such that the set $\mathcal{A}_v$ corresponds to conformal states exactly matching the degeneracies of a single irreducible representation of highest weight $\Delta(v)$. That is, the CDP partition function in the sector with defect number $\ell=2v+1$ is given by 
\be
Z^{\rm CDP}_\ell(q) = Z^{\rm CDP}_{2v+1}(q) = \ch_{\frac{4v^2-1}{8}}(q), \qquad v\in\mathbb{N}_0
\ee
corresponding to a boldfaced term in the expansions given above in (\ref{z0})-(\ref{z2}). In other words, the boldfaced terms comprise the contributions from the admissible configurations.

In the dimer model for $N$ even, as already indicated in (\ref{ZvNR}), {\em all}$\,$ two-column configurations of 
\be
 \mathrm{maximal\ height}:\quad n={\textstyle \frac{N}{2}}\qquad ({\rm Ramond})
\ee 
must be included, even though many of them in general are non-admissible. They nevertheless carry physical information and are accounted for by a combinatorial enumeration reminiscent of the physical combinatorics of the CDP model on the {\em cylinder}~\cite{PRV}, likewise for $N$ even, as outlined in the following.

\begin{figure}
\setlength{\unitlength}{.8pt}
\psset{unit=.8pt}
\begin{center}
\begin{pspicture}(-50,-10)(335,50)
\thicklines
\multirput(0,0)(52,0){7}{\psframe[linewidth=0.0pt,linecolor=lightyellow,fillstyle=solid,
  fillcolor=yellow!40!white](-18,-7.5)(18,40)}
\multirput(0,0)(52,0){7}{\psline[linewidth=0.5pt](-18,45)(-18,-7.5)(18,-7.5)(18,45)}
\multiput(-7,0)(52,0){7}{\multiput(0,0)(0,15){3}{\psarc[linecolor=black,
  linewidth=.5pt,fillstyle=solid,fillcolor=white](0,0){3}{0}{360}}}
\multiput(7,0)(52,0){7}{\multiput(0,0)(0,15){3}{\psarc[linecolor=black,
  linewidth=.5pt,fillstyle=solid,fillcolor=white](0,0){3}{0}{360}}}
\multiput(-7,0)(0,15){3}{\psarc[linecolor=lightred,linewidth=.5pt,fillstyle=solid,
  fillcolor=lightred](0,0){3}{0}{360}}
\multiput(45,0)(0,15){2}{\psarc[linecolor=lightred,linewidth=.5pt,fillstyle=solid,
  fillcolor=lightred](0,0){3}{0}{360}}
\multiput(97,0)(0,15){1}{\psarc[linecolor=lightred,linewidth=.5pt,fillstyle=solid,
  fillcolor=lightred](0,0){3}{0}{360}}
\multiput(215,0)(0,15){1}{\psarc[linecolor=blue,linewidth=.5pt,fillstyle=solid,
  fillcolor=blue](0,0){3}{0}{360}}
\multiput(267,0)(0,15){2}{\psarc[linecolor=blue,linewidth=.5pt,fillstyle=solid,
  fillcolor=blue](0,0){3}{0}{360}}
\multiput(319,0)(0,15){3}{\psarc[linecolor=blue,linewidth=.5pt,fillstyle=solid,
  fillcolor=blue](0,0){3}{0}{360}}
\rput[B](-40,-32){$v$}
\rput[B](0,-32){$-3$}
\rput[B](52,-32){$-2$}
\rput[B](104,-32){$-1$}
\rput[B](156,-32){$0$}
\rput[B](208,-32){$1$}
\rput[B](260,-32){$2$}
\rput[B](312,-32){$3$}
\rput[B](-40,-3){\scriptsize $j=1$}
\rput[B](-40,12){\scriptsize $j=2$}
\rput[B](-40,27){\scriptsize $j=3$}
\end{pspicture}
\end{center}
\smallskip
\caption{Ramond sectors ($N$ even): Two-column configurations with minimal 
energy excitation (\ref{REmin}) for fixed quantum number $v$ 
given by the excess of blue (right) over red (left).
\label{EsigmaR1}}
\end{figure}
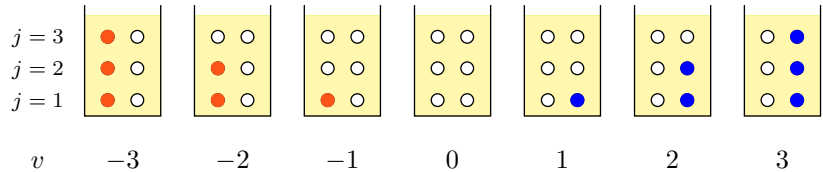

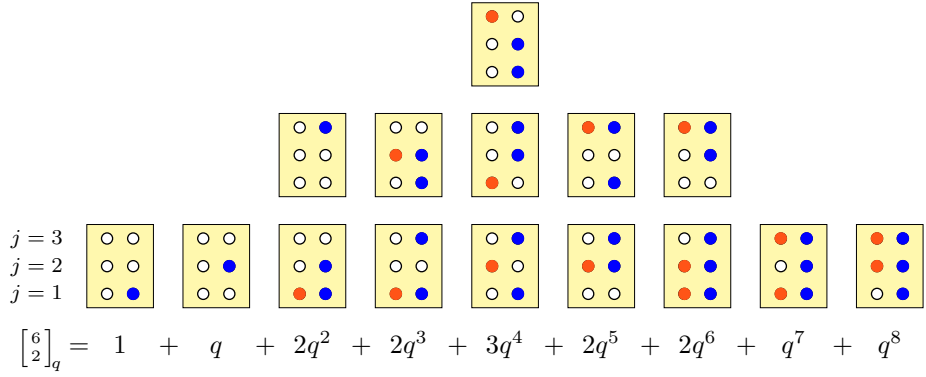
\begin{figure}
\setlength{\unitlength}{.7pt}
\psset{unit=.7pt}
\begin{center}
\begin{pspicture}(-60,-17)(450,170)
\thicklines
\multirput(0,0)(52,0){9}{\psframe[linewidth=0pt,fillstyle=solid,
  fillcolor=yellow!40!white](-18,-7.5)(18,37.5)}
\multirput(104,60)(52,0){5}{\psframe[linewidth=0pt,fillstyle=solid,
  fillcolor=yellow!40!white](-18,-7.5)(18,37.5)}
\multirput(208,120)(52,0){1}{\psframe[linewidth=0pt,fillstyle=solid,
  fillcolor=yellow!40!white](-18,-7.5)(18,37.5)}
\multiput(-7,0)(52,0){9}{\multiput(0,0)(0,15){3}{\psarc[linecolor=black,
  linewidth=.5pt,fillstyle=solid,fillcolor=white](0,0){3}{0}{360}}}
\multiput(7,0)(52,0){9}{\multiput(0,0)(0,15){3}{\psarc[linecolor=black,
  linewidth=.5pt,fillstyle=solid,fillcolor=white](0,0){3}{0}{360}}}
\multiput(97,60)(52,0){5}{\multiput(0,0)(0,15){3}{\psarc[linecolor=black,
  linewidth=.5pt,fillstyle=solid,fillcolor=white](0,0){3}{0}{360}}}
\multiput(111,60)(52,0){5}{\multiput(0,0)(0,15){3}{\psarc[linecolor=black,
  linewidth=.5pt,fillstyle=solid,fillcolor=white](0,0){3}{0}{360}}}
\multiput(201,120)(52,0){1}{\multiput(0,0)(0,15){3}{\psarc[linecolor=black,
  linewidth=.5pt,fillstyle=solid,fillcolor=white](0,0){3}{0}{360}}}
\multiput(215,120)(52,0){1}{\multiput(0,0)(0,15){3}{\psarc[linecolor=black,
  linewidth=.5pt,fillstyle=solid,fillcolor=white](0,0){3}{0}{360}}}
\psarc[linecolor=blue,linewidth=.5pt,fillstyle=solid,fillcolor=blue](7,0){3}{0}{360}
\psarc[linecolor=blue,linewidth=.5pt,fillstyle=solid,fillcolor=blue](59,15){3}{0}{360}
\psarc[linecolor=lightred,linewidth=.5pt,fillstyle=solid,fillcolor=lightred](97,0){3}{0}{360}
\psarc[linecolor=blue,linewidth=.5pt,fillstyle=solid,fillcolor=blue](111,0){3}{0}{360}
\psarc[linecolor=blue,linewidth=.5pt,fillstyle=solid,fillcolor=blue](111,15){3}{0}{360}
\psarc[linecolor=blue,linewidth=.5pt,fillstyle=solid,fillcolor=blue](111,90){3}{0}{360}
\psarc[linecolor=lightred,linewidth=.5pt,fillstyle=solid,fillcolor=lightred](149,0){3}{0}{360}
\psarc[linecolor=lightred,linewidth=.5pt,fillstyle=solid,fillcolor=lightred](149,75){3}{0}{360}
\psarc[linecolor=blue,linewidth=.5pt,fillstyle=solid,fillcolor=blue](163,0){3}{0}{360}
\psarc[linecolor=blue,linewidth=.5pt,fillstyle=solid,fillcolor=blue](163,30){3}{0}{360}
\psarc[linecolor=blue,linewidth=.5pt,fillstyle=solid,fillcolor=blue](163,60){3}{0}{360}
\psarc[linecolor=blue,linewidth=.5pt,fillstyle=solid,fillcolor=blue](163,75){3}{0}{360}
\psarc[linecolor=lightred,linewidth=.5pt,fillstyle=solid,fillcolor=lightred](201,15){3}{0}{360}
\psarc[linecolor=lightred,linewidth=.5pt,fillstyle=solid,fillcolor=lightred](201,60){3}{0}{360}
\psarc[linecolor=lightred,linewidth=.5pt,fillstyle=solid,fillcolor=lightred](201,150){3}{0}{360}
\psarc[linecolor=blue,linewidth=.5pt,fillstyle=solid,fillcolor=blue](215,0){3}{0}{360}
\psarc[linecolor=blue,linewidth=.5pt,fillstyle=solid,fillcolor=blue](215,30){3}{0}{360}
\psarc[linecolor=blue,linewidth=.5pt,fillstyle=solid,fillcolor=blue](215,75){3}{0}{360}
\psarc[linecolor=blue,linewidth=.5pt,fillstyle=solid,fillcolor=blue](215,90){3}{0}{360}
\psarc[linecolor=blue,linewidth=.5pt,fillstyle=solid,fillcolor=blue](215,120){3}{0}{360}
\psarc[linecolor=blue,linewidth=.5pt,fillstyle=solid,fillcolor=blue](215,135){3}{0}{360}
\psarc[linecolor=lightred,linewidth=.5pt,fillstyle=solid,fillcolor=lightred](253,15){3}{0}{360}
\psarc[linecolor=lightred,linewidth=.5pt,fillstyle=solid,fillcolor=lightred](253,90){3}{0}{360}
\psarc[linecolor=blue,linewidth=.5pt,fillstyle=solid,fillcolor=blue](267,15){3}{0}{360}
\psarc[linecolor=blue,linewidth=.5pt,fillstyle=solid,fillcolor=blue](267,30){3}{0}{360}
\psarc[linecolor=blue,linewidth=.5pt,fillstyle=solid,fillcolor=blue](267,60){3}{0}{360}
\psarc[linecolor=blue,linewidth=.5pt,fillstyle=solid,fillcolor=blue](267,90){3}{0}{360}
\psarc[linecolor=lightred,linewidth=.5pt,fillstyle=solid,fillcolor=lightred](305,0){3}{0}{360}
\psarc[linecolor=lightred,linewidth=.5pt,fillstyle=solid,fillcolor=lightred](305,15){3}{0}{360}
\psarc[linecolor=lightred,linewidth=.5pt,fillstyle=solid,fillcolor=lightred](305,90){3}{0}{360}
\psarc[linecolor=blue,linewidth=.5pt,fillstyle=solid,fillcolor=blue](319,0){3}{0}{360}
\psarc[linecolor=blue,linewidth=.5pt,fillstyle=solid,fillcolor=blue](319,15){3}{0}{360}
\psarc[linecolor=blue,linewidth=.5pt,fillstyle=solid,fillcolor=blue](319,30){3}{0}{360}
\psarc[linecolor=blue,linewidth=.5pt,fillstyle=solid,fillcolor=blue](319,75){3}{0}{360}
\psarc[linecolor=blue,linewidth=.5pt,fillstyle=solid,fillcolor=blue](319,90){3}{0}{360}
\psarc[linecolor=lightred,linewidth=.5pt,fillstyle=solid,fillcolor=lightred](357,0){3}{0}{360}
\psarc[linecolor=lightred,linewidth=.5pt,fillstyle=solid,fillcolor=lightred](357,30){3}{0}{360}
\psarc[linecolor=blue,linewidth=.5pt,fillstyle=solid,fillcolor=blue](371,0){3}{0}{360}
\psarc[linecolor=blue,linewidth=.5pt,fillstyle=solid,fillcolor=blue](371,15){3}{0}{360}
\psarc[linecolor=blue,linewidth=.5pt,fillstyle=solid,fillcolor=blue](371,30){3}{0}{360}
\psarc[linecolor=lightred,linewidth=.5pt,fillstyle=solid,fillcolor=lightred](409,15){3}{0}{360}
\psarc[linecolor=lightred,linewidth=.5pt,fillstyle=solid,fillcolor=lightred](409,30){3}{0}{360}
\psarc[linecolor=blue,linewidth=.5pt,fillstyle=solid,fillcolor=blue](423,0){3}{0}{360}
\psarc[linecolor=blue,linewidth=.5pt,fillstyle=solid,fillcolor=blue](423,15){3}{0}{360}
\psarc[linecolor=blue,linewidth=.5pt,fillstyle=solid,fillcolor=blue](423,30){3}{0}{360}
\rput[B](-36,-32){{\scriptsize $\big[{6 \atop 2}\big]_{\!q}\;$}=}
\multiput(21,-32)(52,0){8}{$+$}
\rput[B](0,-32){$1$}
\rput[B](52,-32){$q$}
\rput[B](104,-32){$2q^2$}
\rput[B](156,-32){$2q^3$}
\rput[B](208,-32){$3q^4$}
\rput[B](260,-32){$2q^5$}
\rput[B](312,-32){$2q^6$}
\rput[B](364,-32){$q^7$}
\rput[B](416,-32){$q^8$}
\rput[B](-45,-3){\scriptsize $j=1$}
\rput[B](-45,12){\scriptsize $j=2$}
\rput[B](-45,27){\scriptsize $j=3$}
\end{pspicture}
\end{center}
\smallskip
\caption{Ramond sectors ($N$ even): Combinatorial enumeration by two-column configurations of the $q$-binomial {\scriptsize $\big[{N \atop n-v}\big]_{\!q}$}$=${\scriptsize $\big[{6 \atop 2}\big]_{\!q}$}$=q^{-\frac{1}{2}} \sum q^{\sum_j \delta_j E_j}$. The excess of blue (right) over red (left) is given by the quantum number $v=1$. The lowest energy configuration has energy $E(v)=\half v^2=\half$.   Of the $15$ configurations appearing here, only the top one is non-admissible.
\label{binomR1}}
\end{figure}

The two-column configurations with variation index $v$ are generated combinatorially by starting with the minimum-energy configuration for given $v$ as shown in Figure~\ref{EsigmaR1}. The energy excitation of such a ground state is 
\be
 E(v) = \sum_{j=1}^{|v|} \; (j - \half) = \half v^2,\qquad v\in\mathbb{Z},
\label{REmin}
\ee
while the corresponding conformal weight is
\be
 \Delta(v)=\Delta_0^{{\rm even}}+E(v)=\Delta_{|v|+1,2}\ .
\label{RDmin}
\ee
The generating function for the full excitation spectrum for finite system size $N$ is given by the finitized version of (\ref{genfunceven}),
\be
\prod_{j=1}^n \;  \,\sum_{\ep_j,\mu_j\in\{0,1\}} q^{(\ep_j+\mu_j)(j-{1 \over 2})} \, y^{\mu_j-\ep_j} = q^{n^2 \over 2} \, y^{-n} \: \prod_{k=0}^{N-1} \: (1 + y\,q^{k+{1 \over 2}-n}) = \sum_{v=-n}^n \; \bigg[{N \atop n-v}\bigg]_{\!q} \, q^{v^2 \over 2} \, y^v\,,
\ee
from which it follows that the finite-size generating function for fixed $v$ is incorporated in the $q$-binomial
\be
q^{-E(v)}\sum_{{\mbox{{\scriptsize $v$--two}} \atop \mbox{{\scriptsize columns}}}}q^{\sum_j \delta_jE_j} = \bigg[{N \atop n-v}\bigg]_{\!q}.
\label{qbin}
\ee
Indeed, the elementary excitations (of energy $1$) are generated by either inserting a left-right pair of occupants at level $j=1$ or promoting an occupant at level $j$ to level $j+1$ in the same column. These are the only operations increasing the energy by $1$ while preserving $v$. 
The generation of the corresponding spectrum and its link to (\ref{qbin}) are illustrated in Figure~\ref{binomR1}. We stress that, even though 
\be
 \bigg[{N \atop n-v}\bigg]_{\!q}=\bigg[{N \atop n+v}\bigg]_{\!q}
\label{bineq}
\ee 
as $q$-polynomials, they have {\em different} combinatorial interpretations because they have different quantum numbers: $v$ and $-v$, respectively. The combinatorial interpretations obviously agree in the exceptional case where $v=0$.

This combinatorial enumeration immediately leads to the finitized partition function
\be
 Z_v^{(N)}(q)=q^{-\frac{c}{24}+\Delta_{|v|+1,2}}\bigg[{N \atop n-v}\bigg]_{\!q},\qquad |v|\leq n=\frac{N}{2},
\label{Zvbin}
\ee
and our next objective is to write this in terms of finitized irreducible $c=-2$ characters~\cite{pera,PRV}. 

In the Ramond sector, the relevant such characters are the ones associated with the {\em second} row of the extended Kac table in Figure~\ref{Kac}, and for which the finitization is with respect to an {\em odd} strip in the CDP model. Following~\cite{PRV}, we thus have 
\be
 Z_v^{(N)}(q)=\sum_{r=|v|+1,\,\mathrm{by}\,2}^{n\, \mathrm{or}\, n+1}\ch_{r,2}^{(N+1)}(q),
\label{Zv2n}
\ee
where the finitized character $\ch_{r,s}^{(N+1)}(q)=\ch_{\Delta_{r,s}}^{(N+1)}(q)$ for $s=2$ is given by
\be
 \ch_{r,2}^{(N+1)}(q)=q^{-\frac{c}{24}+\Delta_{r,2}}\frac{1-q^{2r}}{1-q^{N+2}}\bigg[{N+2 \atop n-r+1}\bigg]_{\!q},
\label{chr2}
\ee
and where the upper summation limit depends on the parity of $n-v$. Since $ \ch_{r,2}^{(N+1)}(q)$ vanishes for $r>n+1$, the specification of this upper limit is redundant, as it could be replaced by $\infty$. In the continuum scaling limit $N\to\infty$, we readily recover the decomposition (\ref{vermaeven}),
\be
 \lim_{N\to\infty}Z_v^{(N)}(q)=\sum_{r=|v|+1,\,\mathrm{by}\,2}^\infty\ch_{r,2}(q)=Z_v(q).\qquad \quad(\mbox{Ramond})
\label{limZR}
\ee

It is recalled that only the first term $\ch_{|v|+1,2}(q)$ in (\ref{vermaeven}) and (\ref{limZR}) is related to the CDP model through (\ref{l2v}), as indicated in (\ref{z0})-(\ref{z2}). Likewise for finite system sizes, only some eigenvalues contribute to $\ch_{|v|+1,2}^{(N+1)}(q)$, namely the ones corresponding to {\em admissible} two-column configurations. Here we present a combinatorial prescription which, for given $v$ and $n$, associates {\em every} corresponding two-column configuration with a state in one of the finitized characters appearing in the decomposition (\ref{Zv2n}), giving this decomposition a clear combinatorial content. 

\begin{figure}
\setlength{\unitlength}{.8pt}
\psset{unit=.8pt}
\begin{center}
\begin{pspicture}(0,-8)(108,70)
\thicklines
\multirput(0,0)(102,0){2}{\psframe[linewidth=0pt,fillstyle=solid,
  fillcolor=yellow!40!white](-18,-7.5)(18,37.5)}
\multiput(-7,0)(102,0){2}{\multiput(0,0)(0,15){3}{\psarc[linecolor=black,
  linewidth=.5pt,fillstyle=solid,fillcolor=white](0,0){3}{0}{360}}}
\multiput(7,0)(102,0){2}{\multiput(0,0)(0,15){3}{\psarc[linecolor=black,
  linewidth=.5pt,fillstyle=solid,fillcolor=white](0,0){3}{0}{360}}}
\psarc[linecolor=blue,linewidth=.5pt,fillstyle=solid,fillcolor=blue](7,0){3}{0}{360}
\psarc[linecolor=blue,linewidth=.5pt,fillstyle=solid,fillcolor=blue](7,15){3}{0}{360}
\psarc[linecolor=lightred,linewidth=.5pt,fillstyle=solid,fillcolor=lightred](-7,30){3}{0}{360}
\psarc[linecolor=blue,linewidth=.5pt,fillstyle=solid,fillcolor=blue](109,0){3}{0}{360}
\psarc[linecolor=blue,linewidth=.5pt,fillstyle=solid,fillcolor=blue](109,15){3}{0}{360}
\psarc[linecolor=blue,linewidth=.5pt,fillstyle=solid,fillcolor=blue](109,30){3}{0}{360}
\rput[B](0,-32){$v=1$}
\rput[B](51,11){$\longrightarrow$}
\rput[B](102,-32){$v_{\mathrm{adm}}=3$}
\end{pspicture}
\end{center}
\smallskip
\caption{Illustration of how our prescription turns a non-admissible two-column configuration into an admissible one. The non-admissibility measure of the initial configuration is $t=1$, and its single non-admissibility label is $\kappa_1=3$. The variation index is increased by a multiple of $2$, here from $v=1$ to $v_{\mathrm{adm}}=v+2t=3$. This is the maximal attainable value, cf.$\!\,$ (\ref{UA}). The resulting configuration is the ground-state configuration for $v_{\mathrm{adm}}=3$. In accordance with (\ref{D4D}), its energy excitation satisfies $E(v_{\mathrm{adm}})=E(v)+4$, where $E(v)=E(1)=\frac{1}{2}$.
\label{exn3}}
\end{figure}
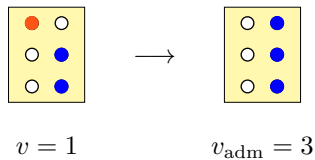

Before giving the general prescription, let us examine $Z_1^{(6)}(q)$ featured in Figure~\ref{binomR1}. From (\ref{Zvbin}), we have
\be
 Z_1^{(6)}(q)=q^{-\frac{c}{24}+\Delta_{2,2}}\bigg[{6 \atop 2}\bigg]_{\!q},
\ee
whereas (\ref{Zv2n}) yields
\be
 Z_1^{(6)}(q)=\ch_{2,2}^{(7)}(q)+\ch_{4,2}^{(7)}(q)=q^{-\frac{c}{24}+\Delta_{2,2}}\big(1+q+2q^2+2q^3+2q^4+2q^5+2q^6+q^7+q^8\big)+q^{-\frac{c}{24}+\Delta_{4,2}}.
\ee
Consistency requires
\be
 \Delta_{2,2}+4=\Delta_{4,2},
\label{D4D}
\ee
which is easily verified. The only two-column configuration in Figure~\ref{binomR1} not contributing to $\ch_{2,2}^{(7)}(q)$ is the single non-admissible configuration appearing at the top. The announced combinatorial prescription turns it into an admissible configuration with $v=3$ contributing to $\ch_{4,2}^{(7)}(q)$ by moving the single occupant in the left column to the right column, as illustrated in Figure~\ref{exn3}. This way of accounting for the sub-leading finitized characters in (\ref{Zv2n}) extends to general $Z_v^{(N)}(q)$ as described in the following.

First, as a measure of `non-admissibility' of a finite two-column configuration, we introduce
\be
 t = \Big|\min_{\kappa\in\mathbb{N}}\{0,\sum_{j\geq\kappa}(\mu_j-\epsilon_j)\}\Big|.
\ee
According to (\ref{adm}), an admissible configuration is thus characterised by a vanishing {\em non-admissibility measure}, $t=0$.

Let $U_v^h$ denote the set of two-column configurations of maximal height $h$ and variation index $v$. The cardinality of this set is
\be
 |U_v^h|=\bigg({2h \atop h-v}\bigg).
\label{U}
\ee
For a configuration in $U_v^h$, the non-admissibility measure $t$ satisfies
\be
 \max\{0,-v\}\leq t\leq t_{\max},\qquad t_{\max}=\lfloor\frac{h-v}{2}\rfloor.
\label{ttt}
\ee

Now, consider a two-column configuration in $U_v^h$ with non-admissibility measure $t$. Associated to this configuration and for every $\tau=1,2,\ldots, t$, we introduce the {\em non-admissibility label} $\kappa_\tau$ as the maximum value of $\kappa\in\{1,2,\ldots,h\}$ for which $\sum_{j\geq\kappa}(\mu_j-\epsilon_j)=-\tau$. That is,
\be
 \kappa_\tau=\max\Big\{\kappa\in\{1,2,\ldots,h\};\ \sum_{j\geq\kappa}(\mu_j-\epsilon_j)=-\tau\Big\}.
\ee
By construction, such a configuration must be of the form indicated to the left in Figure~\ref{kt}. The $t$ occupants at levels $\kappa_\tau$, $\tau=1,2,\ldots,t$, all appear in the left column with unoccupied sites to their right. In addition, the parts of the configuration above the level $\kappa_1$ and in-between the levels $\kappa_{\tau}$ and $\kappa_{\tau+1}$ for every $\tau = 1,2,\ldots,t-1$, make up admissible sub-configurations with zero variation index; the part of the  configuration comprising the lowest $\kappa_t-1$ levels makes up an admissible subconfiguration with variation index $v+t$.

\begin{figure}
\psset{unit=.6cm}
\setlength{\unitlength}{.6cm}
\begin{center}
\begin{pspicture}(0,1)(2,9.5)
\psframe[linewidth=0.0pt,linecolor=yellow!40!white,fillstyle=solid,fillcolor=yellow!40!white](0,0)(2,9)
\psline[linewidth=0.5pt](0,9.5)(0,0)(2,0)(2,9.5)
\psline[linecolor=black,linewidth=0.1pt,linestyle=dashed](-0.2,7.3)(2.2,7.3)
\rput(-0.75,7){$\kappa_1$}
\psarc[linecolor=lightred,linewidth=.5pt,fillstyle=solid,fillcolor=lightred](0.5,7){.12}{0}{360}
\psarc[linecolor=black,linewidth=.5pt,fillstyle=solid,fillcolor=white](1.5,7){.12}{0}{360}
\psline[linecolor=black,linewidth=0.1pt,linestyle=dashed](-0.2,6.7)(2.2,6.7)
\psline[linecolor=black,linewidth=0.1pt,linestyle=dashed](-0.2,5)(2.2,5)
\rput(-0.75,4.7){$\kappa_2$}
\psarc[linecolor=lightred,linewidth=.5pt,fillstyle=solid,fillcolor=lightred](0.5,4.7){.12}{0}{360}
\psarc[linecolor=black,linewidth=.5pt,fillstyle=solid,fillcolor=white](1.5,4.7){.12}{0}{360}
\psline[linecolor=black,linewidth=0.1pt,linestyle=dashed](-0.2,4.4)(2.2,4.4)
\rput(-0.75,3.375){$\vdots$}
\psline[linecolor=black,linewidth=0.1pt,linestyle=dashed](-0.2,2)(2.2,2)
\rput(-0.75,1.7){$\kappa_t$}
\psarc[linecolor=lightred,linewidth=.5pt,fillstyle=solid,fillcolor=lightred](0.5,1.7){.12}{0}{360}
\psarc[linecolor=black,linewidth=.5pt,fillstyle=solid,fillcolor=white](1.5,1.7){.12}{0}{360}
\psline[linecolor=black,linewidth=0.1pt,linestyle=dashed](-0.2,1.4)(2.2,1.4)
\rput(1,3.2){{\scriptsize adm$_{0}$}}
\rput(1,5.85){{\scriptsize adm$_{0}$}}
\rput(1,8.2){{\scriptsize adm$_{0}$}}
\rput(1,0.7){{\scriptsize adm$_{v+t}$}}
\end{pspicture}
\begin{pspicture}(0,1)(4,9.5)
\rput(2,4.7){$\longrightarrow$}
\end{pspicture}
\begin{pspicture}(0,1)(2,9.5)
\psframe[linewidth=0.0pt,linecolor=yellow!40!white,fillstyle=solid,fillcolor=yellow!40!white](0,0)(2,9)
\psline[linewidth=0.5pt](0,9.5)(0,0)(2,0)(2,9.5)
\psline[linecolor=black,linewidth=0.1pt,linestyle=dashed](-0.2,7.3)(2.2,7.3)
\psarc[linecolor=black,linewidth=.5pt,fillstyle=solid,fillcolor=white](0.5,7){.12}{0}{360}
\psarc[linecolor=blue,linewidth=.5pt,fillstyle=solid,fillcolor=blue](1.5,7){.12}{0}{360}
\psline[linecolor=black,linewidth=0.1pt,linestyle=dashed](-0.2,6.7)(2.2,6.7)
\psline[linecolor=black,linewidth=0.1pt,linestyle=dashed](-0.2,5)(2.2,5)
\psarc[linecolor=black,linewidth=.5pt,fillstyle=solid,fillcolor=white](0.5,4.7){.12}{0}{360}
\psarc[linecolor=blue,linewidth=.5pt,fillstyle=solid,fillcolor=blue](1.5,4.7){.12}{0}{360}
\psline[linecolor=black,linewidth=0.1pt,linestyle=dashed](-0.2,4.4)(2.2,4.4)
\psline[linecolor=black,linewidth=0.1pt,linestyle=dashed](-0.2,2)(2.2,2)
\psarc[linecolor=black,linewidth=.5pt,fillstyle=solid,fillcolor=white](0.5,1.7){.12}{0}{360}
\psarc[linecolor=blue,linewidth=.5pt,fillstyle=solid,fillcolor=blue](1.5,1.7){.12}{0}{360}
\psline[linecolor=black,linewidth=0.1pt,linestyle=dashed](-0.2,1.4)(2.2,1.4)
\rput(1,3.2){{\scriptsize adm$_{0}$}}
\rput(1,5.85){{\scriptsize adm$_{0}$}}
\rput(1,8.2){{\scriptsize adm$_{0}$}}
\rput(1,0.7){{\scriptsize adm$_{v+t}$}}
\end{pspicture}
\end{center}
\caption{\label{kt} Prescription for turning a non-admissible two-column configuration into an admissible one. The initial configuration has non-admissibility labels $\kappa_1,\,\kappa_2,\,\ldots,\,\kappa_t$ and variation index $v$, while the resulting admissible configuration has variation index $v_{\mathrm{adm}}=v+2t$.}
\end{figure}

Our prescription to associate an admissible configuration to a non-admissible configuration of the previous type is simply to move the $t$ occupants at levels $\kappa_1,\kappa_2,\ldots,\kappa_t$ from the left to the right column (as in Figure~\ref{kt} and illustrated in Figure~\ref{exn3}), resulting in a configuration with variation index
\be
 v_{\mathrm{adm}}=v+2t.
\label{vadm}
\ee

It remains to be demonstrated that the admissible configurations obtained in this way exactly correspond to the ones associated with the finitized characters appearing in the decomposition (\ref{Zv2n}). We do this by establishing that the prescription yields a bijective map 
\be
 U_v^n\ \to\ \bigcup_{v_{\mathrm{adm}}=|v|,\,\mbox{{\scriptsize by}}\,2}^{n-1\, \mbox{{\scriptsize or}}\, n}A_{v_{\mathrm{adm}}}^{(n)},
\label{UA}
\ee
where the upper limit depends on the parity of $n-v$. The bijectivity of this prescription map follows by observing (i) that the cardinality of the union in (\ref{UA}) equals the one in (\ref{U}) for $h=n$; and (ii) that every admissible two-column configuration in $A_{v_{\mathrm{adm}}}^{(n)}$ with variation index written as $v_{\mathrm{adm}}=v+2t$, where $t,v+t\geq0$, is of the form indicated to the right in Figure~\ref{kt}.

The range of the (disjoint) union index in (\ref{UA}) is dictated by the possible values for $v_{\mathrm{adm}}$ following from (\ref{ttt}) and (\ref{vadm}). These values are readily seen to be the ones labelling the summation in the decomposition (\ref{Zv2n}), reflecting that the admissible configurations in $A_{v_{\mathrm{adm}}}^{(n)}$ are the ones contributing to the finitized character $\ch_{v_{\mathrm{adm}}+1,2}^{(N+1)}(q)$ in (\ref{Zv2n}). Each of the finitized characters in (\ref{Zv2n}) therefore appears as the partition function for a fixed value of $v$ {\it and} a fixed non-admissibility measure $t \geq 0$,
\be
Z_{v;t}^{(N)}(q) = {\rm ch}^{(N+1)}_{|v|+1+2t,2}(q)\,.
\ee
It is noted that the generalised Catalan number 
\be
 \ch_{|v|+1+2t,2}^{(N+1)}(1)=\bigg({N \atop n-|v|-2t}\bigg)-\bigg({N \atop n-|v|-2t-2}\bigg),
\label{Catalan2}
\ee
defined by evaluating the corresponding finitized character at $q=1$, gives the number of two-column configurations with variation index $v$ and non-admissibility measure $t$.

The finitized {\em full} partition function is obtained by ignoring the separation into $v$-sectors, and is given by
\be
 Z^{(N)}(q)=\sum_{v=-n}^nZ_v^{(N)}(q)=\sum_{r=1}^{n+1}r\,\ch_{r,2}^{(N+1)}(q).
\ee
This shows that the finitized character $\ch_{r,2}^{(N+1)}(q)$ appears exactly $r$ times in the corresponding full partition function provided $r\leq n+1$, and thus explains the noted appearance (once, twice and three times) of the irreducible characters of conformal weights $\Delta_{1,2}=-\frac{1}{8}$, $\Delta_{2,2}=\frac{3}{8}$ and $\Delta_{2,3}=\frac{15}{8}$ in (\ref{z0})-(\ref{z2}).

We conclude this analysis of the Ramond sector by reconsidering the full partition function in the continuum scaling limit, that is,
\be
 Z(q)=\lim_{N\to\infty}Z^{(N)}(q)=\sum_{v\in\mathbb{Z}}Z_v(q)=\sum_{r\in\mathbb{N}}r\,\ch_{r,2}(q).
\label{ZsumR}
\ee
In terms of the ${\cal W}$-irreducible characters~\cite{Flohr96,Kausch9510}
\be
 \hat\chi_{-\frac{1}{8}}^{}(q)=\frac{1}{\eta(q)}\vartheta_{0,2}(q)=\sum_{r\in2\mathbb{N}-1}r\,\ch_{r,2}(q),\qquad \hat\chi_{\frac{3}{8}}^{}(q)=\frac{1}{\eta(q)}\vartheta_{2,2}(q)=\sum_{r\in2\mathbb{N}}r\,\ch_{r,2}(q),
\label{WR}
\ee
the partition function (\ref{ZsumR}) thus reads
\be
 Z(q)=\hat\chi_{-\frac{1}{8}}(q)+\hat\chi_{\frac{3}{8}}(q).
\ee
This is in accordance with (\ref{zeven}) since $\theta_3(q)=\vartheta_{0,2}(q)+\vartheta_{2,2}(q)$.

\subsection{Physical combinatorics in the Neveu-Schwarz sector: \boldmath{$N$} odd}
\label{SecPhysNS}

It is recalled that, for $N$ odd, the variation index $v$ takes one of the $N+1$ values
\be
 v\in[-n,n]\cap\big(\mathbb{Z}-\half\big),\qquad n={\textstyle \frac{N}{2}}\in\mathbb{N}-\half.
\ee

From (\ref{specodd}), with $c=-2$ and $\Delta_0^{{\rm odd}}=0$, the conformal energy levels for $N$ odd follow from
\be
 E=E_0+{\alpha\pi \over 2N} \sum_{j \geq 1} \; 2j \:\delta_j = {\alpha\pi \over N}\Big\{-{c \over 24} + \sum_{j \geq 1} \; j \:\delta_j\Big\}\,,
\label{enodd}
\ee
where $\delta_j=0,1,1$ or 2. No contribution to the value of $v$ is made if $\delta_j=0,2$, while $\delta_j=1$ contributes $+1$ or $-1$ to $v$. In addition, every energy level is doubly degenerate; one of the paired energy levels gets an extra contribution to $v$ equal to $+{1 \over 2}$, the other an extra contribution equal to $-{1 \over 2}$. For finite excitations, we thus introduce the {\em excess parameter} 
\be
 w= \sum_{j=1}^{n-\frac{1}{2}} (\mu_j - \ep_j),
\label{w}
\ee
so that 
\be
 w = v \pm \half,\qquad v = w \mp \half,\qquad w\in\{-n+\half,\,-n+{\textstyle \frac{3}{2}},\,\ldots,\,n-\half\}\subset\mathbb{Z}.
\label{wv}
\ee 
Since $|w|\leq n-\half$, we have the unique relation $w=v-\half$ (resp.$\!$\, $w=v+\half$) if $v=n$ (resp.$\!$\, $v=-n$). For $N$ even, there was no need to introduce a separate excess parameter since the variation index $v$ itself was sufficient to distinguish between degenerate energy levels, and we simply had (\ref{vme}). As discussed in the following, the excess parameter plays a key role in the physical combinatorics for $N$ odd. 

As we did for $N$ even, let us write 
\be
 \delta_j = \ep_j + \mu_j,\qquad \ep_j,\mu_j\in\{0,1\},\qquad j\in\mathbb{N}.
\ee
To every eigenvalue (\ref{enodd}) characterised by $\{\ep_j,\mu_j\}_{j \geq 1}$, we again associate a two-column configuration with occupants in the left or right column at levels $j$ for which $\ep_j=1$ or $\mu_j=1$, respectively. For finite excitations, the maximal height of these two-column configurations is
\be
 \mathrm{maximal\ height}:\quad n-{\textstyle \frac{1}{2}}={\textstyle \frac{N-1}{2}},\qquad ({\rm Neveu-Schwarz})
\ee 
and the excess of occupants in the right column over the left column is given by the excess parameter $w$. The notion of admissibility and the non-admissibility measure and labels are as for $N$ even.

From (\ref{vermaodd}), the partition functions of the first few $v$-sectors read
\bea
&& Z_{-\frac{1}{2}}(q) = \mbox{\boldmath${\ch_{0}}(q)$} + \ch_{1}(q) + \ch_{3}(q)+\ch_{6}(q)+\ch_{10}(q)+\ch_{15}(q)+ \ldots, \label{z-1h}\\
\noalign{\medskip}
&& Z_{\frac{1}{2}}(q) = \mbox{\boldmath$\ch_{0}(q) + \ch_{1}(q)$} + \ch_{3}(q) +\ch_{6}(q) + \ch_{10}(q) + \ch_{15}(q) + \ldots, \label{z1h}\\
\noalign{\medskip}
&& Z_{\frac{3}{2}}(q) = \mbox{\boldmath$\ch_{1}(q)+\ch_3(q)$} +\ch_{6}(q) + \ch_{10}(q) + \ch_{15}(q) + \ch_{21}(q) + \ldots, \quad \ Z_{-\frac{3}{2}}(q) = Z_{\frac{3}{2}}(q),\qquad \label{z3h}\\
\noalign{\medskip}
&& Z_{\frac{5}{2}}(q) = \mbox{\boldmath$\ch_{3}(q)+\ch_6(q)$} + \ch_{10}(q) + \ch_{15}(q) + \ch_{21}(q) + \ch_{28}(q) + \ldots, \quad Z_{-\frac{5}{2}}(q) = Z_{\frac{5}{2}}(q).\qquad \label{z5h}
\eea
The rationale for writing some of these characters in boldface is as above and is recapitulated below. 

For finite system size $N$ and variation index $v$, a natural finitization of the partition function $Z_v(q)$ is given by
\be
 Z_v^{(N)}(q)=q^{-\frac{c}{24}+\Delta_0^{\mathrm{odd}}}\sum_{{\mbox{{\scriptsize $v$--two}} \atop \mbox{{\scriptsize columns}}}}q^{\sum_j \delta_jE_j},
\label{ZvNNS}
\ee
where the sum is over all two-column configurations of maximal height $n-\half$ and excess parameter $w=v\pm\half$, and where the elementary excitation energy of an occupant at level $j$ follows from (\ref{enodd}) and is given by
\be
 E_j=j.
\label{EjNS}
\ee

\begin{figure}
\setlength{\unitlength}{.8pt}
\psset{unit=.8pt}
\begin{center}
\begin{pspicture}(-50,-20)(340,50)
\thicklines
\multirput(0,0)(52,0){7}{\psframe[linewidth=0.0pt,linecolor=lightyellow,fillstyle=solid,
  fillcolor=yellow!40!white](-18,-7.5)(18,40)}
\multirput(0,0)(52,0){7}{\psline[linewidth=0.5pt](-18,45)(-18,-7.5)(18,-7.5)(18,45)}
\multiput(-7,0)(52,0){7}{\multiput(0,0)(0,15){3}{\psarc[linecolor=black,
  linewidth=.5pt,fillstyle=solid,fillcolor=white](0,0){3}{0}{360}}}
\multiput(7,0)(52,0){7}{\multiput(0,0)(0,15){3}{\psarc[linecolor=black,
  linewidth=.5pt,fillstyle=solid,fillcolor=white](0,0){3}{0}{360}}}
\multiput(-7,0)(0,15){3}{\psarc[linecolor=red,linewidth=.5pt,fillstyle=solid,
  fillcolor=red](0,0){3}{0}{360}}
\multiput(45,0)(0,15){2}{\psarc[linecolor=red,linewidth=.5pt,fillstyle=solid,
  fillcolor=red](0,0){3}{0}{360}}
\multiput(97,0)(0,15){1}{\psarc[linecolor=red,linewidth=.5pt,fillstyle=solid,
  fillcolor=red](0,0){3}{0}{360}}
\multiput(215,0)(0,15){1}{\psarc[linecolor=blue,linewidth=.5pt,fillstyle=solid,
  fillcolor=blue](0,0){3}{0}{360}}
\multiput(267,0)(0,15){2}{\psarc[linecolor=blue,linewidth=.5pt,fillstyle=solid,
  fillcolor=blue](0,0){3}{0}{360}}
\multiput(319,0)(0,15){3}{\psarc[linecolor=blue,linewidth=.5pt,fillstyle=solid,
  fillcolor=blue](0,0){3}{0}{360}}
\rput[B](-40,-50){$v$}
\rput[B](0,-50){$-\frac{7}{2}$}
\rput[B](52,-50){$-\frac{5}{2}$}
\rput[B](104,-50){$-\frac{3}{2}$}
\rput[B](156,-50){$\pm\frac{1}{2}$}
\rput[B](208,-50){$\frac{3}{2}$}
\rput[B](260,-50){$\frac{5}{2}$}
\rput[B](312,-50){$\frac{7}{2}$}
\rput[B](-40,-28){$w$}
\rput[B](0,-28){$-3$}
\rput[B](52,-28){$-2$}
\rput[B](104,-28){$-1$}
\rput[B](156,-28){$0$}
\rput[B](208,-28){$1$}
\rput[B](260,-28){$2$}
\rput[B](312,-28){$3$}
\rput[B](-40,-3){\scriptsize $j=1$}
\rput[B](-40,12){\scriptsize $j=2$}
\rput[B](-40,27){\scriptsize $j=3$}
\end{pspicture}
\end{center}
\smallskip
\caption{Neveu-Schwarz sectors ($N$ odd): Two-column configurations with minimal 
energy excitation (\ref{NSEmin}) for fixed quantum number $v$. The excess of blue (right) over red (left) is given by the excess parameter $w$. For these minimal energy configurations, $w$ is related to the variation index $v$ by $w=v-\frac{1}{2}$ if $v>0$ and $w=v+\frac{1}{2}$ if $v<0$.
\label{EsigmaNS}}
\end{figure}
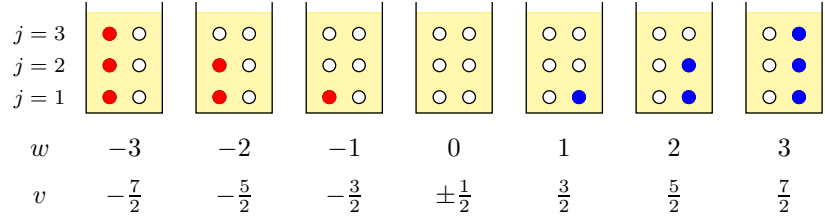

The two-column configurations with variation index $v$ are generated combinatorially by starting with the minimum-energy configuration for given $v$ as shown in Figure~\ref{EsigmaNS}. The energy excitation of such a ground state is 
\be
 E(v)=\half v^2-{\textstyle {1 \over 8}},\qquad v\in\mathbb{Z}-{\textstyle {1 \over 2}},
\label{NSEmin}
\ee
while the corresponding conformal weight is
\be
 \Delta(v)=\Delta_0^{{\rm odd}}+E(v)=\Delta_{|v|+\frac{1}{2},1}\ .
\label{NSDmin}
\ee

The finitized version of (\ref{genfuncodd}) yields the generating function of the full spectrum for finite system size,
\be
(y^{\frac{1}{2}} + y^{-\frac{1}{2}}) \prod_{j=1}^{n-\frac{1}{2}}\; \sum_{\ep_j,\mu_j\in\{0,1\}} q^{(\ep_j+\mu_j)j} \, y^{\mu_j-\ep_j} = q^{{n^2 \over 2}-{1 \over 8}} y^{-n} \prod_{k=0}^{N-1} (1 + y\,q^{k+{1 \over 2}-n}) = \sum_{v=-n}^n \bigg[{N \atop n-v}\bigg]_{\!q} \, q^{{v^2 \over 2}-{1 \over 8}} \, y^v,
\ee
and immediately leads to the finitized partition function
\be
 Z_v^{(N)}(q)=q^{-\frac{c}{24}+\Delta_{|v|+\frac{1}{2},1}}\bigg[{N \atop n-v}\bigg]_{\!q},\qquad |v|\leq n.
\label{ZvbinNS}
\ee

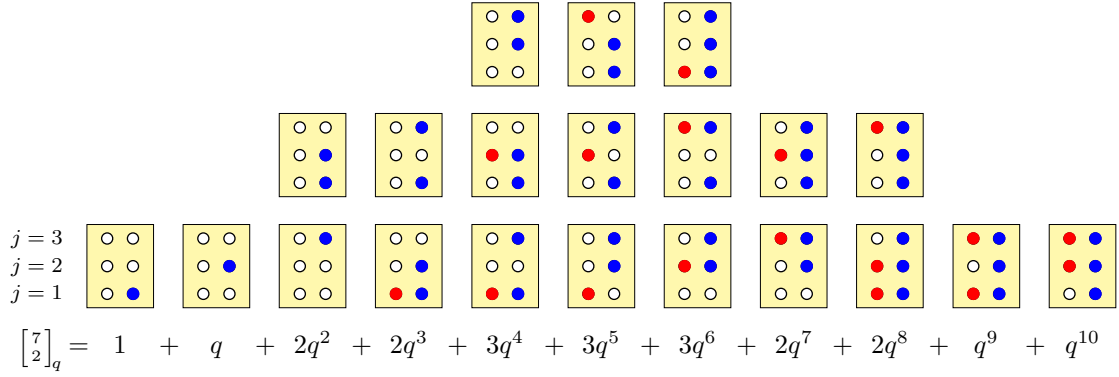
\begin{figure}
\setlength{\unitlength}{.7pt}
\psset{unit=.7pt}
\begin{center}
\begin{pspicture}(-60,-17)(545,170)
\thicklines
\multirput(0,0)(52,0){11}{\psframe[linewidth=0pt,fillstyle=solid,
  fillcolor=yellow!40!white](-18,-7.5)(18,37.5)}
\multirput(104,60)(52,0){7}{\psframe[linewidth=0pt,fillstyle=solid,
  fillcolor=yellow!40!white](-18,-7.5)(18,37.5)}
\multirput(208,120)(52,0){3}{\psframe[linewidth=0pt,fillstyle=solid,
  fillcolor=yellow!40!white](-18,-7.5)(18,37.5)}
\multiput(-7,0)(52,0){11}{\multiput(0,0)(0,15){3}{\psarc[linecolor=black,
  linewidth=.5pt,fillstyle=solid,fillcolor=white](0,0){3}{0}{360}}}
\multiput(7,0)(52,0){11}{\multiput(0,0)(0,15){3}{\psarc[linecolor=black,
  linewidth=.5pt,fillstyle=solid,fillcolor=white](0,0){3}{0}{360}}}
\multiput(97,60)(52,0){7}{\multiput(0,0)(0,15){3}{\psarc[linecolor=black,
  linewidth=.5pt,fillstyle=solid,fillcolor=white](0,0){3}{0}{360}}}
\multiput(111,60)(52,0){7}{\multiput(0,0)(0,15){3}{\psarc[linecolor=black,
  linewidth=.5pt,fillstyle=solid,fillcolor=white](0,0){3}{0}{360}}}
\multiput(201,120)(52,0){3}{\multiput(0,0)(0,15){3}{\psarc[linecolor=black,
  linewidth=.5pt,fillstyle=solid,fillcolor=white](0,0){3}{0}{360}}}
\multiput(215,120)(52,0){3}{\multiput(0,0)(0,15){3}{\psarc[linecolor=black,
  linewidth=.5pt,fillstyle=solid,fillcolor=white](0,0){3}{0}{360}}}
\psarc[linecolor=blue,linewidth=.5pt,fillstyle=solid,fillcolor=blue](7,0){3}{0}{360}
\psarc[linecolor=blue,linewidth=.5pt,fillstyle=solid,fillcolor=blue](59,15){3}{0}{360}
\psarc[linecolor=blue,linewidth=.5pt,fillstyle=solid,fillcolor=blue](111,30){3}{0}{360}
\psarc[linecolor=blue,linewidth=.5pt,fillstyle=solid,fillcolor=blue](111,60){3}{0}{360}
\psarc[linecolor=blue,linewidth=.5pt,fillstyle=solid,fillcolor=blue](111,75){3}{0}{360}
\psarc[linecolor=red,linewidth=.5pt,fillstyle=solid,fillcolor=red](149,0){3}{0}{360}
\psarc[linecolor=blue,linewidth=.5pt,fillstyle=solid,fillcolor=blue](163,0){3}{0}{360}
\psarc[linecolor=blue,linewidth=.5pt,fillstyle=solid,fillcolor=blue](163,15){3}{0}{360}
\psarc[linecolor=blue,linewidth=.5pt,fillstyle=solid,fillcolor=blue](163,60){3}{0}{360}
\psarc[linecolor=blue,linewidth=.5pt,fillstyle=solid,fillcolor=blue](163,90){3}{0}{360}
\psarc[linecolor=red,linewidth=.5pt,fillstyle=solid,fillcolor=red](201,0){3}{0}{360}
\psarc[linecolor=red,linewidth=.5pt,fillstyle=solid,fillcolor=red](201,75){3}{0}{360}
\psarc[linecolor=blue,linewidth=.5pt,fillstyle=solid,fillcolor=blue](215,0){3}{0}{360}
\psarc[linecolor=blue,linewidth=.5pt,fillstyle=solid,fillcolor=blue](215,30){3}{0}{360}
\psarc[linecolor=blue,linewidth=.5pt,fillstyle=solid,fillcolor=blue](215,60){3}{0}{360}
\psarc[linecolor=blue,linewidth=.5pt,fillstyle=solid,fillcolor=blue](215,75){3}{0}{360}
\psarc[linecolor=blue,linewidth=.5pt,fillstyle=solid,fillcolor=blue](215,135){3}{0}{360}
\psarc[linecolor=blue,linewidth=.5pt,fillstyle=solid,fillcolor=blue](215,150){3}{0}{360}
\psarc[linecolor=red,linewidth=.5pt,fillstyle=solid,fillcolor=red](253,0){3}{0}{360}
\psarc[linecolor=red,linewidth=.5pt,fillstyle=solid,fillcolor=red](253,75){3}{0}{360}
\psarc[linecolor=red,linewidth=.5pt,fillstyle=solid,fillcolor=red](253,150){3}{0}{360}
\psarc[linecolor=blue,linewidth=.5pt,fillstyle=solid,fillcolor=blue](267,15){3}{0}{360}
\psarc[linecolor=blue,linewidth=.5pt,fillstyle=solid,fillcolor=blue](267,30){3}{0}{360}
\psarc[linecolor=blue,linewidth=.5pt,fillstyle=solid,fillcolor=blue](267,60){3}{0}{360}
\psarc[linecolor=blue,linewidth=.5pt,fillstyle=solid,fillcolor=blue](267,90){3}{0}{360}
\psarc[linecolor=blue,linewidth=.5pt,fillstyle=solid,fillcolor=blue](267,120){3}{0}{360}
\psarc[linecolor=blue,linewidth=.5pt,fillstyle=solid,fillcolor=blue](267,135){3}{0}{360}
\psarc[linecolor=red,linewidth=.5pt,fillstyle=solid,fillcolor=red](305,15){3}{0}{360}
\psarc[linecolor=red,linewidth=.5pt,fillstyle=solid,fillcolor=red](305,90){3}{0}{360}
\psarc[linecolor=red,linewidth=.5pt,fillstyle=solid,fillcolor=red](305,120){3}{0}{360}
\psarc[linecolor=blue,linewidth=.5pt,fillstyle=solid,fillcolor=blue](319,15){3}{0}{360}
\psarc[linecolor=blue,linewidth=.5pt,fillstyle=solid,fillcolor=blue](319,30){3}{0}{360}
\psarc[linecolor=blue,linewidth=.5pt,fillstyle=solid,fillcolor=blue](319,60){3}{0}{360}
\psarc[linecolor=blue,linewidth=.5pt,fillstyle=solid,fillcolor=blue](319,90){3}{0}{360}
\psarc[linecolor=blue,linewidth=.5pt,fillstyle=solid,fillcolor=blue](319,120){3}{0}{360}
\psarc[linecolor=blue,linewidth=.5pt,fillstyle=solid,fillcolor=blue](319,135){3}{0}{360}
\psarc[linecolor=blue,linewidth=.5pt,fillstyle=solid,fillcolor=blue](319,150){3}{0}{360}
\psarc[linecolor=red,linewidth=.5pt,fillstyle=solid,fillcolor=red](357,30){3}{0}{360}
\psarc[linecolor=red,linewidth=.5pt,fillstyle=solid,fillcolor=red](357,75){3}{0}{360}
\psarc[linecolor=blue,linewidth=.5pt,fillstyle=solid,fillcolor=blue](371,15){3}{0}{360}
\psarc[linecolor=blue,linewidth=.5pt,fillstyle=solid,fillcolor=blue](371,30){3}{0}{360}
\psarc[linecolor=blue,linewidth=.5pt,fillstyle=solid,fillcolor=blue](371,60){3}{0}{360}
\psarc[linecolor=blue,linewidth=.5pt,fillstyle=solid,fillcolor=blue](371,75){3}{0}{360}
\psarc[linecolor=blue,linewidth=.5pt,fillstyle=solid,fillcolor=blue](371,90){3}{0}{360}
\psarc[linecolor=red,linewidth=.5pt,fillstyle=solid,fillcolor=red](409,0){3}{0}{360}
\psarc[linecolor=red,linewidth=.5pt,fillstyle=solid,fillcolor=red](409,15){3}{0}{360}
\psarc[linecolor=red,linewidth=.5pt,fillstyle=solid,fillcolor=red](409,90){3}{0}{360}
\psarc[linecolor=blue,linewidth=.5pt,fillstyle=solid,fillcolor=blue](423,0){3}{0}{360}
\psarc[linecolor=blue,linewidth=.5pt,fillstyle=solid,fillcolor=blue](423,15){3}{0}{360}
\psarc[linecolor=blue,linewidth=.5pt,fillstyle=solid,fillcolor=blue](423,30){3}{0}{360}
\psarc[linecolor=blue,linewidth=.5pt,fillstyle=solid,fillcolor=blue](423,60){3}{0}{360}
\psarc[linecolor=blue,linewidth=.5pt,fillstyle=solid,fillcolor=blue](423,75){3}{0}{360}
\psarc[linecolor=blue,linewidth=.5pt,fillstyle=solid,fillcolor=blue](423,90){3}{0}{360}
\psarc[linecolor=red,linewidth=.5pt,fillstyle=solid,fillcolor=red](461,0){3}{0}{360}
\psarc[linecolor=red,linewidth=.5pt,fillstyle=solid,fillcolor=red](461,30){3}{0}{360}
\psarc[linecolor=blue,linewidth=.5pt,fillstyle=solid,fillcolor=blue](475,0){3}{0}{360}
\psarc[linecolor=blue,linewidth=.5pt,fillstyle=solid,fillcolor=blue](475,15){3}{0}{360}
\psarc[linecolor=blue,linewidth=.5pt,fillstyle=solid,fillcolor=blue](475,30){3}{0}{360}
\psarc[linecolor=red,linewidth=.5pt,fillstyle=solid,fillcolor=red](513,15){3}{0}{360}
\psarc[linecolor=red,linewidth=.5pt,fillstyle=solid,fillcolor=red](513,30){3}{0}{360}
\psarc[linecolor=blue,linewidth=.5pt,fillstyle=solid,fillcolor=blue](527,0){3}{0}{360}
\psarc[linecolor=blue,linewidth=.5pt,fillstyle=solid,fillcolor=blue](527,15){3}{0}{360}
\psarc[linecolor=blue,linewidth=.5pt,fillstyle=solid,fillcolor=blue](527,30){3}{0}{360}
\rput[B](-36,-32){{\scriptsize $\big[{7 \atop 2}\big]_{\!q}\;$}=}
\multiput(21,-32)(52,0){10}{$+$}
\rput[B](0,-32){$1$}
\rput[B](52,-32){$q$}
\rput[B](104,-32){$2q^2$}
\rput[B](156,-32){$2q^3$}
\rput[B](208,-32){$3q^4$}
\rput[B](260,-32){$3q^5$}
\rput[B](312,-32){$3q^6$}
\rput[B](364,-32){$2q^7$}
\rput[B](416,-32){$2q^8$}
\rput[B](468,-32){$q^9$}
\rput[B](520,-32){$q^{10}$}
\rput[B](-45,-3){\scriptsize $j=1$}
\rput[B](-45,12){\scriptsize $j=2$}
\rput[B](-45,27){\scriptsize $j=3$}
\end{pspicture}
\end{center}
\smallskip
\caption{Neveu-Schwarz sectors ($N$ odd): Combinatorial enumeration by two-column configurations of the $q$-binomial {\scriptsize $\big[{N \atop m}\big]_{\!q}$}$=${\scriptsize $\big[{7 \atop 2}\big]_{\!q}$}$=q^{-1} \sum q^{\sum_j \delta_j E_j}$. The excess of blue (right) over red (left) is given by the excess parameter $w=1$ or $w=2$ related to the quantum number $v=n-m=\frac{3}{2}$ through $w=v\pm\frac{1}{2}$. The lowest energy configuration has energy $E(v)=\half v^2-\frac{1}{8}=1$. Of the $21$ configurations appearing here, only the top one in the middle column is non-admissible. The six configurations appearing at the top of the 3rd, 4th, 5th, 7th, 8th and 9th columns have $w=2$; the remaining 15 configurations have $w=1$.
\label{binomNS}}
\end{figure}

In this case, the elementary excitations (of energy $1$) are generated by either inserting a left or right occupant at level $j=1$, provided the excess parameter is $w=v+\half$ or $w=v-\half$, respectively, or promoting an occupant at level $j$ to level $j+1$ in the same column. These are the only operations increasing the energy by $1$ while preserving $v$. The generation of the corresponding spectrum is illustrated in Figure~\ref{binomNS}. 

Following~\cite{PRV}, we then have
\be
 Z_v^{(N)}(q)=\sum_{r=|v|+\frac{1}{2}}^{n+\frac{1}{2}}\ch_{r,1}^{(N+1)}(q)=\sum_{w=v\pm\frac{1}{2}}\Big( \sum_{r=|w|+1,\,\mathrm{by}\,2}^{n-\frac{1}{2}\,\mathrm{or}\,n+\frac{1}{2}}\ch_{r,1}^{(N+1)}(q)\Big),
\label{Zv2nNS}
\ee
where the finitized characters are given by
\be
 \ch_{r,1}^{(N+1)}(q)=q^{-\frac{c}{24}+\Delta_{r,1}}\frac{1-q^{r}}{1-q^{n+\frac{1}{2}}}\bigg[{N+1 \atop n-r+\frac{1}{2}}\bigg]_{\!q}.
\label{chr1}
\ee
The leading character in (\ref{Zv2nNS}) appears for $r=|v|+\half$. In the continuum scaling limit $N\to\infty$, we readily recover the decomposition (\ref{vermaodd}),
\be
 \lim_{N\to\infty}Z_v^{(N)}(q)=\sum_{r=|v|+\frac{1}{2}}^\infty\ch_{r,1}(q)=Z_v(q).\qquad \quad(\mbox{Neveu-Schwarz})
\label{limZNS}
\ee

Admissibility of a two-column configuration requires $t=0$ and hence $w\geq0$, so with the CDP defect number $\ell\geq0$ related to the variation index $v$ by $\ell=2v+1$, the characters also present in the CDP model, as $\ell$ varies over the even non-negative integers, are the ones in boldface in (\ref{z-1h})-(\ref{z5h}). The appearance of two characters in boldface for $\ell>0$ reflects that the corresponding CDP representation is {\em reducible}~\cite{pera,rasm}. The decomposition of the associated set of admissible configurations of maximal height $h$ reads
\be
 \mathcal{A}_w^{(h)}=\bigcup_{m=0}^{h-w} \mathcal{A}^{(h)}_{m,m+w}.
\ee 
In somewhat sloppy notation, we thus have
\be
Z_{-\frac{1}{2}}^{(N)} \supset \mathcal{A}_0^{(n-\frac{1}{2})}\,, \qquad Z_{v}^{(N)} \supset \Big(\mathcal{A}_{v-\frac{1}{2}}^{(n-\frac{1}{2})} \cup \mathcal{A}_{v+\frac{1}{2}}^{(n-\frac{1}{2})}\Big), \qquad v\in\mathbb{N}-\half,
\ee
whose continuum scaling limit is indicated by
\be
Z_{-\frac{1}{2}} \supset \mathcal{A}_0\,, \qquad Z_{v} \supset \Big(\mathcal{A}_{v-\frac{1}{2}} \cup \mathcal{A}_{v+\frac{1}{2}}\Big), \qquad v\in\mathbb{N}-\half.
\ee

It is noted that the generalised Catalan numbers defined by evaluating the finitized characters (\ref{chr1}) at $q=1$,
\be
 \ch_{r,1}^{(N+1)}(1)=\bigg({N-1 \atop n-r+\frac{1}{2}}\bigg)-\bigg({N-1 \atop n-r-\frac{3}{2}}\bigg),
\label{Catalan1}
\ee
verify that
\be
 \sum_{r=|v|+\frac{1}{2}}^{n+\frac{1}{2}}\ch_{r,1}^{(N+1)}(1)=\bigg({N \atop n-v}\bigg)=Z_v^{(N)}(1).
\label{Z2}
\ee
Separating this sum into the two corresponding $w$ streams as in the double sum in (\ref{Zv2nNS}), we have the refined identities
\be
 \sum_{r=|v-\frac{1}{2}|+1,\,\mathrm{by}\,2}^{n-\frac{1}{2}\,\mathrm{or}\,n+\frac{1}{2}}\ch_{r,1}^{(N+1)}(1)=\bigg({N-1 \atop n-|v-\frac{1}{2}|-\frac{1}{2}}\bigg),\qquad \sum_{r=|v+\frac{1}{2}|+1,\,\mathrm{by}\,2}^{n+\frac{1}{2}\,\mathrm{or}\,n-\frac{1}{2}}\ch_{r,1}^{(N+1)}(1)=\bigg({N-1 \atop n-|v+\frac{1}{2}|-\frac{1}{2}}\bigg).
\ee

Our combinatorial prescription for associating non-admissible configurations with physical states in the finitized characters and partition functions extends from the Ramond sector described in Section~\ref{SecRamond} to the Neveu-Schwarz sector discussed here. A priori, there appears to be a caveat, though, as moving occupants from the left to the right column increases the excess parameter $w$ by a multiple of $2$, 
\be
 w_{\mathrm{adm}}=w+2t,
\ee
but not necessarily the variation index $v=w\pm\half$. However, it is exactly an increment in the excess parameter which ensures that the bijective map (\ref{UA}) in the Ramond sector extends to a bijective map in the Neveu-Schwarz sector,
\be
 \Big(U_{v-\frac{1}{2}}^{n-\frac{1}{2}}\cup U_{v+\frac{1}{2}}^{n-\frac{1}{2}}\Big)\ \to\ \bigcup_{w_{\mathrm{adm}}=|v|-\frac{1}{2}}^{n-\frac{1}{2}}\mathcal{A}_{w_{\mathrm{adm}}}^{(n)}.
\label{UANS}
\ee

Let us examine $Z_{\frac{3}{2}}^{(7)}(q)$ featured in Figure~\ref{binomNS}. From (\ref{ZvbinNS}), we have
\be
 Z_{\frac{3}{2}}^{(7)}(q)=q^{-\frac{c}{24}+\Delta_{2,1}}\bigg[{7 \atop 2}\bigg]_{\!q},
\ee
whereas (\ref{Zv2nNS}) yields
\bea
 Z_{\frac{3}{2}}^{(7)}(q)&=&=\ch_{2,1}^{(8)}(q)+\ch_{3,1}^{(8)}(q)+\ch_{4,1}^{(8)}(q)\nonumber\\
 &=&q^{-\frac{c}{24}+\Delta_{2,1}}\big(1+q+q^2+q^3+2q^4+2q^5+2q^6+q^7+q^8+q^9+q^{10}\big)\nonumber\\
 &+&q^{-\frac{c}{24}+\Delta_{3,1}}\big(1+q+q^2+q^4+q^5+q^6\big)+q^{-\frac{c}{24}+\Delta_{4,1}}.
\eea
Consistency requires the easily verified relations
\be
 \Delta_{2,1}+2=\Delta_{3,1},\qquad \Delta_{2,1}+5=\Delta_{4,1}.
\ee
With reference to the $21$ configurations in Figure~\ref{binomNS}, the 14 configurations contributing to $\ch_{2,1}^{(8)}(q)$ are the admissible ones with $w=1$. The six configurations with $w=2$, which are all admissible, give rise to $\ch_{3,1}^{(8)}(q)$, while the single non-admissible configuration has $w=1$. According to our combinatorial prescription (\ref{UANS}), it contributes to $\ch_{4,1}^{(8)}(q)$ after moving the single occupant in the left column (at level $3$) to the right. Since $\Delta_{2,1}=1$, $\Delta_{3,1}=3$ and $\Delta_{4,1}=6$, this is in accordance with the decomposition and use of boldface in (\ref{z3h}).

We conclude this analysis of the Neveu-Schwarz sector by ignoring the separation into $v$-sectors. The corresponding finitized full partition function is given by
\be
 Z^{(N)}(q)=\sum_{v=-n}^nZ_v^{(N)}(q)=\sum_{r=1}^{n+\frac{1}{2}}2r\,\ch_{r,1}(q),
\ee
whose continuum scaling limit yields the full partition function,
\be
 Z(q)=\lim_{N\to\infty}Z^{(N)}(q)=\sum_{v\in\mathbb{Z}+\frac{1}{2}}Z_v(q)=2\sum_{r\in\mathbb{N}}r\,\ch_{r,1}(q).
\label{ZsumNS}
\ee
In terms of the ${\cal W}$-irreducible characters~\cite{Flohr96,Kausch9510}
\be
 \hat\chi_{0}^{}(q)=\frac{1}{2\eta(q)}\Big(\vartheta_{1,2}(q)+\eta^3(q)\Big)=\sum_{r\in2\mathbb{N}-1}r\,\ch_{r,1}(q),\qquad
 \hat\chi_{1}^{}(q)=\frac{1}{2\eta(q)}\Big(\vartheta_{1,2}(q)-\eta^3(q)\Big)=\sum_{r\in2\mathbb{N}}r\,\ch_{r,1}(q),
\label{WNS}
\ee
we have
\be
 \hat\chi_0^{}(q)+\hat\chi_1^{}(q)=\frac{1}{\eta(q)}\vartheta_{1,2}(q)=\sum_{r=1}^\infty r\,\ch_{r,1}(q)
\ee
and hence
\be
 Z(q)=2[\hat\chi_0^{}(q)+\hat\chi_1^{}(q)].
\ee
This is in accordance with (\ref{zodd}) since $\theta_2(q)=2\vartheta_{1,2}(q)$.

\section{Periodic boundary conditions}
\label{SecPeriodic}

In the periodic case, the diagonalisation of $T^2$ was carried out in~\cite{lieb}. The relation of the intermediate fermions $C^{}_j,C^\dagger_j$ to the Pauli matrices is the same as in the open case, see (\ref{jw}). The passage to the fermions $\eta_k^{},\eta^\dagger_k$ is slightly different, however, and distinguishes the sectors with $v$ even from the sectors with $v$ odd. As the notion of sectors is essentially trivial for $N$ odd, our considerations will be split according to the parities of $N$ and $v$. For $N$ even, the eigenvectors of $T^2$ have even (resp.$\!$\, odd) fermion number in the $v$ even (resp.$\!$\, odd) sector. Because of similarities with the results for open boundary conditions, we refer to these sectors as the Ramond ($v$ even) and Neveu-Schwarz ($v$ odd) sectors.

\subsection{Ramond sector: \boldmath{$N$} even, \boldmath{$v$} even}

For $N$ and $v$ even, in terms of the fermions $\eta^{}_k,\eta_k^\dagger$, $1 \leq k \leq N$, the square of the transfer matrix factorizes as \cite{lieb}
\be
T^2 =\bigotimes_{k=1}^{{N \over 2}}\: A_k, \quad  A_k = \exp\Big\{2\a\sin{q^{}_k} \; \eta^{}_k \, \eta^{}_{N+1-k}\Big\} \, \exp\Big\{2\a\sin{q^{}_k} \; \eta^\dagger_{N+1-k} \, \eta^\dagger_k\Big\}, \quad q_k = {(2k-1)\pi \over N}.
\ee
The blocks $A_k$ are all four-dimensional and easily diagonalised, and one obtains the relevant spectrum of $T^2$ by retaining the eigenvectors with an {\em even} fermion number \cite{lieb}. 

Since $\sin{q^{}_k} = \sin{q_{{N \over 2}+1-k}}$, we see that $A^{}_k = A_{{N \over 2}+1-k}$, implying that $T^2$ is the product of two almost identical factors
\be
T^2 = \Big[\bigotimes_{k=1}^{\lfloor{N+2 \over 4}\rfloor}\: A_k\Big] \otimes \Big[\bigotimes_{k=1}^{\lfloor{N \over 4}\rfloor}\: A_k\Big].
\label{factor}
\ee
The first, left, factor in square brackets has a spectrum given by
\be 
\lambda^{\rm R}_l = \prod_{k=1}^{\lfloor{N+2 \over 4}\rfloor} \: \Big[\sqrt{1+\a^2 \sin^2{q_k}} + \a \sin{q_k}\Big]^{2(1-\ep_k - \mu_k)}\,, \qquad \quad q_k = {(2k-1)\pi \over N}\,,
\label{evenl}
\ee 
while the spectrum of the second, right, factor is
\be 
\lambda^{\rm R}_r = \prod_{k=1}^{\lfloor{N \over 4}\rfloor} \: \Big[\sqrt{1+\a^2 \sin^2{q_k}} + \a \sin{q_k}\Big]^{2(1-\bar\ep_k - \bar\mu_k)}\,, \qquad \quad q_k = {(2k-1)\pi \over N}\,,
\label{evenr}
\ee 
where the numbers $\ep_k,\mu_k,\bar\ep_k,\bar\mu_k$ take the values 0,1 as before. We find that the  energy levels have the following asymptotic form
\be
E^{\rm R}_l = N'f_{{\rm bulk}}+{\alpha\pi \over N'}\Big\{-{c \over 24}-\frac{1}{8} + \sum_{j \geq 1} (j-\half) \delta_j\Big\}+\ldots,\;\;
E^{\rm R}_r = N'f_{{\rm bulk}}+{\alpha\pi \over N'}\Big\{-{c \over 24}-\frac{1}{8} + \sum_{j \geq 1} (j-\half) \bar\delta_j\Big\}+\ldots
\ee
with $\delta_j = \ep_j+\mu_j$, $\bar\delta_j = \bar\ep_j + \bar\mu_j$ and $N'={N \over 2}$. In particular, we see that the conformal spectra of the two factors in $T^2$ are identical, and are recognised as two separate Ramond sectors (\ref{eneven}) of {\em even} system size $N'$, but with vanishing boundary free energy, $f_{{\rm bdy}}^{\rm R}=0$ (as expected for periodic boundary conditions). 

As $T^2$ is the tensor product of the two factors, it follows that the total energy levels $E^{\rm R}$ can be written in terms of a `left' and a `right' contribution, 
\be
 E^{\rm R} = E^{\rm R}_l + E^{\rm R}_r,
\label{ER}
\ee 
in a way reminiscent of {\em non-chiral} conformal representations. The two `chiral halves' correspond to two open Ramond sectors with $q$ replaced by $q'=q^2$,
\be
 q=e^{-\frac{\alpha\pi M}{N}}\ \to\ q'=e^{-\frac{\alpha\pi M}{N'}}=e^{-\frac{2\alpha\pi M}{N}}=q^2.
\ee

The variation index $v$ is given as an eigenvalue of the operator
\bea
\V = \sum_{j=1}^N \,(-1)^j \, C^\dagger_j C^{\phantom\dagger}_j &=&
 \eta^\dagger_{\frac{N+2}{4}} \, \eta^{}_{\frac{3N+2}{4}} + \eta^\dagger_{\frac{3N+2}{4}} \eta^{}_{\frac{N+2}{4}} \nonumber\\
&+&\!\sum_{1 \leq k \leq \frac{N}{4}} (\eta^\dagger_k \, \eta^{}_{\frac{N}{2}+k} + \eta^\dagger_{\frac{N}{2}+k} \, \eta^{\phantom{\dagger}}_{k} + \eta^\dagger_{\frac{N}{2}+1-k} \, \eta^{\phantom{\dagger}}_{N+1-k} + \eta^\dagger_{N+1-k} \, \eta^{}_{\frac{N}{2}+1-k}),
\label{VR}
\eea
and is an {\em even} integer in the range
\be
 v\in\{-{\textstyle \frac{N}{2}},\,-{\textstyle \frac{N}{2}}+1,\,\ldots,\,{\textstyle \frac{N}{2}}\}.
\label{vset}
\ee 
The first two $\eta^\dagger\eta$ terms in (\ref{VR}) are present only for $N=2 \bmod 4$.

The way the eigenvalues $v$ of $\V$ are computed respects the left-right factorisation just given, allowing us to write
\be
 v=v_l+v_r,
\label{vvv}
\ee
where
\be
 v_l = \sum_j (\mu_j-\ep_j),\qquad v_r = \sum_j (\bar\mu_j-\bar\ep_j).
\label{vsum}
\ee 
Since the variation index $v$ is even, the two chiral parts are not independent, as we have the {\em gluing condition}
\be
 v_l+v_r\in2\mathbb{Z}.\qquad ({\rm Ramond})
\label{vv2R}
\ee
Ignoring this gluing condition, by counting all conformal states, gives rise to a generating function equal to the square of that of the open case for even system sizes, namely $\theta^2_3(y|q')/\eta^2(q')$ where $q'=q^2$. The physical partition function in the Ramond sector follows by  imposing the Ramond gluing condition, and it is therefore obtained by retaining only the terms with {\em even} $y$-powers in the previous function, that is,
\be
 Z^{\rm R}(q';y) = {\theta_3^2(y|q') + \theta_4^2(y|q') \over 2\eta^2(q')}, \qquad q' = e^{-\frac{2\alpha\pi M}{N}}.
\label{ZRy}
\ee

To determine the contributions from the various $v$-sectors, $Z^{\rm R}_v(q')$, where
\be
 Z^{\rm R}(q';y)=\sum_{v\in2\mathbb{Z}}Z^{\rm R}_v(q')y^v,
\ee
it is convenient to first consider the finitization,
\be
 Z^{{\rm R}(N)}(q';y)=\sum_{v=-2\lfloor\frac{N}{4}\rfloor,\,{\rm by}\,2}^{2\lfloor\frac{N}{4}\rfloor}Z_v^{{\rm R}(N)}(q')y^v,
\ee
 of this expansion following from the separation into chiral halves. To account for all states for a finite system size $N$, we need to treat the two chiral halves (slightly) differently for $N=2$ mod $4$, as follows from the (conventional) form (\ref{factor}). We therefore consider that $h_l\geq h_r$, where
\be
 h_l=\lfloor{\textstyle \frac{N+2}{4}}\rfloor,\qquad h_r=\lfloor{\textstyle \frac{N}{4}}\rfloor,
\ee
are the maximal heights of the two-column configurations in the left and right chiral halves, respectively. We thus have
\bea
 Z_v^{{\rm R}(N)}(q')&=&\sum_{{v_l,v_r\in\mathbb{Z}\atop v_l+v_r=v}}Z_{v_l}^{(2\lfloor\frac{N+2}{4}\rfloor)}(q')Z_{v_r}^{(2\lfloor\frac{N}{4}\rfloor)}(q')\nonumber\\
 &=&\sum_{{v_l,v_r\in\mathbb{Z}\atop v_l+v_r=v}}(q')^{-\frac{c}{24}+\Delta_{|v_l|+1,2}}\bigg[{2\lfloor\frac{N+2}{4}\rfloor \atop \lfloor\frac{N+2}{4}\rfloor-v_l}\bigg]_{\!q'} (q')^{-\frac{c}{24}+\Delta_{|v_r|+1,2}}\bigg[{2\lfloor\frac{N}{4}\rfloor \atop \lfloor\frac{N}{4}\rfloor-v_r}\bigg]_{\!q'}\nonumber\\
 &=&\sum_{r=1}^{\lfloor\frac{N+6}{4}\rfloor}\sum_{\bar r=1}^{\lfloor\frac{N+4}{4}\rfloor}Z_{v;r,\bar r}^{\rm R}\,\ch_{r,2}^{(2\lfloor\frac{N+2}{4}\rfloor+1)}(q')\,\ch_{\bar r,2}^{(2\lfloor\frac{N}{4}\rfloor+1)}(q'),
\eea
where
\be
 Z_{v;r,\bar r}^{\rm R}=\frac{1}{4}(1+(-1)^{r+\bar r})\Big(\max\{|v|,r+\bar r\}-\max\{|v|,|r-\bar r|\}\Big).
\ee
This has a strong resemblance to the corresponding finitization of the Ramond sector of the CDP model on the cylinder~\cite{PRV}, but differs crucially by including also {\em negative} variation indices $v$ as opposed to only non-negative defect numbers $\ell$ in the CDP model. A similar difference between the dimer model and the CDP model appears in the corresponding Neveu-Schwarz sectors, see (\ref{ZvNSN}). As a consequence, the sum of the partition functions in the Ramond and Neveu-Schwarz sectors in the dimer model is {\em modular invariant}, as discussed below, whereas the similar sum in the CDP model is {\em not} modular invariant, as discussed in~\cite{PRV}. 

In the continuum scaling limit, we see that the partition function in the Ramond sector with variation index $v$ is given by
\be
 Z_v^{\rm R}(q')=\sum_{r,\bar r\in\mathbb{N}}Z_{v;r,\bar r}^{\rm R}\,\ch_{r,2}(q')\,\ch_{\bar r,2}(q').
\ee
Ignoring the separation into $v$-sectors yields the total partition function in the Ramond sector, obtained by setting $y=1$,
\be
  Z^{{\rm R}(N)}(q') = Z^{{\rm R}(N)}(q';1) = \sum_{r=1}^{\lfloor\frac{N+6}{4}\rfloor}\sum_{\bar r=1}^{\lfloor\frac{N+4}{4}\rfloor}Z_{r,\bar r}^{\rm R}\,\ch_{r,2}^{(2\lfloor\frac{N+2}{4}\rfloor+1)}(q')\,\ch_{\bar r,2}^{(2\lfloor\frac{N}{4}\rfloor+1)}(q')
\ee
for finite $N$, and
\be 
 Z^{\rm R}(q') = Z^{{\rm R}}(q';1) = \sum_{r,\bar r\in\mathbb{N}}Z_{r,\bar r}^{\rm R}\,\ch_{r,2}(q')\,\ch_{\bar r,2}(q'),
\label{ZR}
\ee
in the continuum scaling limit, where
\be
 Z_{r,\bar r}^{\rm R}=\sum_{v\in2\mathbb{Z}}Z_{v;r,\bar r}^{\rm R}=\frac{1}{2}(1+(-1)^{r+\bar r})r\bar r.
\ee
In terms of ${\cal W}$-irreducible characters (\ref{WR}), we thus have
\be
 Z^{\rm R}(q')=\hat\chi_{-\frac{1}{8}}^{2}(q')+\hat\chi_{\frac{3}{8}}^{2}(q').
\label{ZRhat}
\ee
This is in accordance with (\ref{ZRy}) since $\theta_3(q')=\vartheta_{0,2}(q')+\vartheta_{2,2}(q')$ and $\theta_4(q')=\vartheta_{0,2}(q')-\vartheta_{2,2}(q')$.

\subsection{Neveu-Schwarz sector: \boldmath{$N$} even, \boldmath{$v$} odd}

For $N$ even and $v$ odd, in terms of the fermions $\eta^{}_k,\eta_k^\dagger$, $0 \leq k \leq N-1$, the square of the transfer matrix factorizes as
\be
T^2 = (A_0 \otimes A_{\frac{N}{2}}) \otimes\bigotimes_{k=1}^{{N \over 2}-1}\: A_k, \qquad  A_k = \exp\Big\{2\a\sin{q^{}_k} \; \eta^{}_k \, \eta^{}_{N-k}\Big\} \, \exp\Big\{2\a\sin{q^{}_k} \; \eta^\dagger_{N-k} \, \eta^\dagger_k\Big\}, \quad q_k = {2\pi k \over N}.
\ee
The blocks $A_k$ are four-dimensional, while $A_0=\mathbb{I}_2$ and $A_{\frac{N}{2}}=\mathbb{I}_2$ act in the spaces spanned by $\{|0\ra,\, \eta^\dagger_0|0\ra\}$ and $\{|0\ra,\, \eta^\dagger_{N \over 2}|0\ra\}$, respectively. As for $v$ even, $T^2$ can be written as
\be
T^2 = \Big[A_0 \otimes \bigotimes_{k=1}^{\lfloor{N \over 4}\rfloor}\: A_k\Big] \otimes \Big[A_{N \over 2} \otimes \bigotimes_{k=1}^{\lfloor{N-2 \over 4}\rfloor}\: A_k\Big].
\label{factorodd}
\ee

These two blocks are easily diagonalised, with eigenvalues $\lambda_l^{\rm NS}$, $\lambda_r^{\rm NS}$ given by formulae similar to the even case, (\ref{evenl}) and (\ref{evenr}). For $N$ large, we find that the energies are given by
\be
E^{\rm NS}_l = N'f_{{\rm bulk}}+{\alpha\pi \over N'}\Big\{-{c \over 24} + \sum_{j \geq 1} \; j \:\delta_j\Big\} + \ldots \,,\quad
E^{\rm NS}_r = N'f_{{\rm bulk}}+{\alpha\pi \over N'}\Big\{-{c \over 24} + \sum_{j \geq 1} \; j \:\bar\delta_j\Big\} + \ldots
\ee
with $\delta_j = \ep_j+\mu_j$, $\bar\delta_j = \bar\ep_j + \bar\mu_j$ and $N'={N \over 2}$. Each  `chiral' energy level is doubly degenerate, because of $A_0=\bi_2$ and $A_{N \over 2}=\bi_2$ appearing in the two blocks. Therefore, the two chiral halves each have exactly the same conformal pattern (\ref{enodd}) as in the open case of {\em odd} system size $N'$, but with vanishing boundary free energy, $f_{{\rm bdy}}^{\rm NS}=0$. 

It follows that the full energies $E^{\rm NS}$ separate into two chiral contributions, 
\be
 E^{\rm NS} = E^{\rm NS}_l + E^{\rm NS}_r,
\label{ENS}
\ee 
every such energy level being counted four times. They do not form the relevant spectrum in this sector however, as we have to retain the eigenvectors with an {\em odd} fermion number, that is with an {\em odd} variation index. 

The variation index $v$ is given as an eigenvalue of the operator
\bea
\V = \sum_{j=1}^N \,(-1)^j \, C^\dagger_j C^{\phantom\dagger}_j &=&
\eta^\dagger_0 \, \eta^{}_{{N\over 2}} + \eta^\dagger_{{N\over2}} \eta^{}_0+\eta^\dagger_{{N\over4}} \, \eta^{}_{{3N\over4}} + \eta^\dagger_{{3N\over4}} \eta^{}_{{N\over4}} \nonumber\\
&+&\!\sum_{1 \leq k < {N\over4}} (\eta^\dagger_k \, \eta^{}_{{N\over2}+k} + \eta^\dagger_{{N\over2}+k} \, \eta^{\phantom{\dagger}}_{k} + \eta^\dagger_{{N\over2}-k} \, \eta^{\phantom{\dagger}}_{N-k} + \eta^\dagger_{N-k} \, \eta^{}_{{N\over2}-k}),
\eea
and must be restricted to odd integer values between $-{N\over2}$ and ${N\over2}$. The third and fourth $\eta^\dagger\eta$ terms are present only if $N=0 \bmod 4$.

As in the Ramond sector, the variation index respects the separation into chiral halves and is therefore written as
\be
 v=v_l+v_r, \qquad v_l = \sum_j (\mu_j-\ep_j),\qquad v_r = \sum_j (\bar\mu_j-\bar\ep_j),
\ee
although the attained values are different. In particular, one eigenstate in the block $A_0 \otimes A_{\frac{N}{2}}$ contributes an additional $+1$ to $v$, another one contributes $-1$, and the last two contribute 0, so that all together, they account for a factor $(y + y^{-1}+ 2) = (y^{\frac{1}{2}} + y^{-\frac{1}{2}})^2$ in the generating function. In the Neveu-Schwarz sector, the gluing condition is
\be
 v=v_l+v_r\in2\mathbb{Z}+1.\qquad ({\rm Neveu-Schwarz})
\ee
Ignoring this gluing condition, by counting all conformal states, gives rise to a generating function equal to the square of that of the open case for odd system sizes, namely $\theta^2_2(y|q')/\eta^2(q')$ where $q'=q^2$. The physical partition function in the Neveu-Schwarz sector follows by  imposing the Neveu-Schwarz gluing condition, and it is therefore obtained by retaining only the terms with {\em odd} $y$-powers in the previous function, that is,
\be
Z^{\rm NS}(q';y) = {\theta_2^2(y|q') - \theta_1^2(y|q') \over 2\eta^2(q')}, \qquad q' = e^{-\frac{2\alpha\pi M}{N}}.
\label{ZNSy}
\ee

The contributions from the various $v$-sectors, $Z^{\rm NS}_v(q')$, follow from
\be
 Z^{\rm NS}(q';y)=\sum_{v\in2\mathbb{Z}+1}Z^{\rm NS}_v(q')y^v,
\ee
whose finitization reads
\be
 Z^{{\rm NS}(N)}(q';y)=\sum_{v=-2\lfloor\frac{N+2}{4}\rfloor+1,\,{\rm by}\,2}^{2\lfloor\frac{N+2}{4}\rfloor-1}Z_v^{{\rm NS}(N)}(q')y^v.
\ee
The finitized partition function for fixed $v$ is given by
\bea
 Z_v^{{\rm NS}(N)}(q')&=&\sum_{{v_l,v_r\in\mathbb{Z}+\frac{1}{2}\atop v_l+v_r=v}}Z_{v_l}^{(2\lfloor\frac{N}{4}\rfloor+1)}(q')Z_{v_r}^{(2\lfloor\frac{N-2}{4}\rfloor+1)}(q')\nonumber\\
 &=&\sum_{{v_l,v_r\in\mathbb{Z}+\frac{1}{2}\atop v_l+v_r=v}}(q')^{-\frac{c}{24}+\Delta_{|v_l|+\frac{1}{2},1}}\bigg[{2\lfloor\frac{N}{4}\rfloor+1 \atop \lfloor\frac{N}{4}\rfloor+\frac{1}{2}-v_l}\bigg]_{\!q'} (q')^{-\frac{c}{24}+\Delta_{|v_r|+\frac{1}{2},1}}\bigg[{2\lfloor\frac{N-2}{4}\rfloor+1 \atop \lfloor\frac{N-2}{4}\rfloor+\frac{1}{2}-v_r}\bigg]_{\!q'}\nonumber\\
 &=&\sum_{r=1}^{\lfloor\frac{N+4}{4}\rfloor}\sum_{\bar r=1}^{\lfloor\frac{N+2}{4}\rfloor}Z_{v;r,\bar r}^{\rm NS}\,\ch_{r,1}^{(2\lfloor\frac{N+4}{4}\rfloor)}(q')\,\ch_{\bar r,1}^{(2\lfloor\frac{N+2}{4}\rfloor)}(q'),
\label{ZvNSN}
\eea
where
\be
 Z_{v;r,\bar r}^{\rm NS}=\max\{|v|,r+\bar r\}-\max\{|v|,|r-\bar r|\}.
\ee
In the continuum scaling limit, we thus have
\be
 Z_v^{\rm NS}(q')=\sum_{r,\bar r\in\mathbb{N}}Z_{v;r,\bar r}^{\rm NS}\,\ch_{r,1}(q')\,\ch_{\bar r,1}(q').
\ee

Ignoring the separation into $v$-sectors yields
\be
 Z^{{\rm NS}(N)}(q')=Z^{{\rm NS}(N)}(q';1)=\sum_{r=1}^{\lfloor\frac{N+4}{4}\rfloor}\sum_{\bar r=1}^{\lfloor\frac{N+2}{4}\rfloor}Z_{r,\bar r}^{\rm NS}\,\ch_{r,1}^{(2\lfloor\frac{N+4}{4}\rfloor)}(q')\,\ch_{\bar r,1}^{(2\lfloor\frac{N+2}{4}\rfloor)}(q')
\ee
and
\be 
 Z^{\rm NS}(q')=Z^{{\rm NS}}(q';1)=\sum_{r,\bar r\in\mathbb{N}}Z_{r,\bar r}^{\rm NS}\,\ch_{r,1}(q')\,\ch_{\bar r,1}(q'),
\label{ZNS}
\ee
where
\be
 Z_{r,\bar r}^{\rm NS}=\sum_{v\in2\mathbb{Z}+1}Z_{v;r,\bar r}^{\rm NS}=2r\bar r.
\ee
In terms of the ${\cal W}$-irreducible characters (\ref{WNS}), we thus have
\be
 Z^{\rm NS}(q')=2[\hat\chi_0^{}(q')+\hat\chi_1^{}(q')]^2.
\ee
This is in accordance with (\ref{ZNSy}) since $\theta_1(1|q')=0$ and $\theta_2(q')=2\vartheta_{1,2}(q')$.

\subsection{Full partition function and modular invariance for \boldmath{$N$} even}

The full physical partition function for $N$ even is obtained by adding the contributions from the Ramond and Neveu-Schwarz sectors, 
\be
Z(q';y) = Z^{{\rm R}}(q';y)+Z^{{\rm NS}}(q';y)={-\theta_1^2(y|q') + \theta_2^2(y|q') + \theta_3^2(y|q') + \theta_4^2(y|q') \over 2\eta^2(q')}, \qquad q' = e^{-\frac{2\alpha\pi M}{N}}.
\ee
This partition function may be expanded in powers of $y$ to give the partition functions for individual sectors. However, this merely yields the decompositions already discussed above.

We can relax our considerations by not keeping track of the sectors. This corresponds to setting $y=1$. In this case, we readily recover the torus partition function 
\be
Z(q') =Z(q';1) =Z^{\rm R}(q')+Z^{\rm NS}(q')= {\theta_2^2(q') + \theta_3^2(q') + \theta_4^2(q') \over 2 \eta^2(q')}
\ee
originally obtained by Ferdinand~\cite{ferd}. In terms of ${\cal W}$-irreducible characters, this partition function reads
\be
 Z(q')=\hat\chi_{-\frac{1}{8}}^{2}(q')+\hat\chi_{\frac{3}{8}}^{2}(q')+2[\hat\chi_0^{}(q')+\hat\chi_1^{}(q')]^2.
\ee
This is modular invariant and also appears as the partition function for symplectic fermions as described by the triplet model~\cite{gaka}.

\subsection{Periodic boundary conditions for \boldmath{$N$} odd}

The full spectrum for $N$ odd is the same with periodic boundary conditions as it is with open boundary conditions, where the open spectrum is given in (\ref{enodd}). The patterns of degeneracies, in particular, are thus identical. In the periodic case, each eigenvalue of the form (\ref{enodd}) indeed occurs twice; once in the even fermionic sector, and once in the odd fermionic sector. It follows that the partial partition functions corresponding to an even or odd fermionic number, respectively, are equal, and that the full partition function is the same in the odd open and odd periodic cases. We recall that this partition function is given by (\ref{zodd}). 

As discussed in Section~\ref{Section:SectorsSpectra}, the even and odd subspaces coincide with the two invariant sectors,  $\mathcal{E}_+$ and $\mathcal{E}_-$, defined in (\ref{Eplus}). Since the two partial partition functions are equal, we need not worry about which is which, and accordingly, we write the partition function as
\be
 Z(q)=Z_+(q)+Z_-(q),
\ee
where
\be
 Z_+(q) = Z_-(q) = \frac{\theta_2(q)}{2\eta(q)} = \frac{\vartheta_{1,2}(q)}{\eta(q)}=\hat\chi_0^{}(q)+\hat\chi_1^{}(q),\qquad q=e^{-\frac{\alpha\pi M}{N}}.
\ee
The corresponding finitizations follow readily from the finitizations in the open case discussed in Section~\ref{SecPhysNS}.

\section{Concluding remarks}
\label{SecConclusion}

We have examined the conformal properties of the dimer model based on the working hypothesis that the model is described by a conformal field theory with central charge $c=-2$. Our analysis is significantly finer than previously carried out in the literature, and this improvement was made possible by our introduction of a quantum number called the variation index. This quantum number is present for finite lattice systems and partitions the set of dimer configurations into sectors. Finitizations of the corresponding conformal spectra are described combinatorially in a way reminiscent of similar studies~\cite{pera,PRV} of the critical dense polymer model. Since our results are based on an exact determination of the transfer matrix eigenvalues in the various sectors, they are analytical in nature.

There are still many open problems associated with the dimer model. We hope to return elsewhere with (i) a detailed study of finite-size corrections; and (ii) a discussion of the possibility of having non-trivial Jordan blocks arising in the continuum scaling limit. Both of these studies are expected to help settling the long-standing conundrum with the central charge in the dimer model. We stress in this regard that, even though the results of the present work are consistent with a conformal field theory with $c=-2$, they are not definitive.

\subsection*{Acknowledgments}

\noindent
JR is supported by the Australian Research Council under the Future Fellowship scheme, project number FT100100774, and PR acknowledges the support of the Belgian Interuniversity Attraction Poles Program P6/02. PR is a Senior Research Associate of the Belgian National Fund for Scientific Research (FNRS). The authors thank Jesper L.$\!\,$ Jacobsen, Alexi Morin-Duchesne and Paul A.$\!\,$ Pearce for useful discussions and comments, and Nickolay Izmailian for help with the references.

\appendix

\section{Theta functions}
\label{AppTheta}

\def\numberbysection{\@addtoreset{equation}{section}
     \def\theequation{\thesection.\arabic{equation}}}
\numberbysection

We use the following convention for Jacobi's theta functions,
\bea
\theta_1(y|q)&\!=\!& -{\rm i} \sum_{r \in \Z+{1 \over 2}} \; (-1)^{r-{1 \over 2}} \, y^r \, q^{r^2 \over 2} =\, -{\rm i} \, \sqrt{y} \, q^{{1\over 8}} \, \prod_{n=1}^\infty \,(1-q^n)(1 - y q^{n})(1 - y^{-1} q^{n-1}), 
\\[.1cm]
\theta_2(y|q)&\!=\!& \sum_{r \in \Z+{1 \over 2}} \; y^r \, q^{r^2 \over 2} =\, \sqrt{y} \, q^{{1\over 8}} \, \prod_{n=1}^\infty \,(1-q^n)(1 + y q^{n})(1 + y^{-1} q^{n-1}), 
\\[.1cm]
\theta_3(y|q) &\!=\!& \sum_{n \in \Z} \; y^n \, q^{n^2 \over2} = \,\prod_{n=1}^\infty \,(1-q^n)(1 + y q^{n-\frac{1}{2}})(1 + y^{-1} q^{n-\frac{1}{2}}), 
\\[.1cm]
\theta_4(y|q) &\!=\!& \sum_{n \in \Z} \; (-1)^n \, y^n \, q^{n^2 \over 2} = \,\prod_{n=1}^\infty \,(1-q^n)(1 - y q^{n-\frac{1}{2}})(1 - y^{-1} q^{n-\frac{1}{2}}),
\eea
here written in terms of the nomes $y$ and $q$. At $y=1$, they reduce to the standard theta functions, 
\be
 \theta_j(q)=\theta_j(1|q),\qquad j=1,2,3,4,
\ee
where $\theta_1(q)=0$ and
\be
 \theta_2(q)=\sum_{n\in\mathbb{Z}}q^{\frac{1}{2}(n+\frac{1}{2})^2},\qquad
 \theta_3(q)=\sum_{n\in\mathbb{Z}}q^{\frac{n^2}{2}},\qquad
 \theta_4(q)=\sum_{n\in\mathbb{Z}}(-1)^nq^{\frac{n^2}{2}}.
\label{thetaq}
\ee
The related Dedekind eta function is defined by
\be
 \eta(q)=q^\frac{1}{24}\prod_{n=1}^\infty(1-q^n).
\label{eta}
\ee
In terms of the generalised theta functions
\be
 \vartheta_{n,m}(q)=\sum_{k\in\mathbb{Z}}q^{m(k+\frac{n}{2m})^2},
\label{vartheta}
\ee
we have
\be
 \theta_2(q)=2\vartheta_{1,2}(q),\qquad\theta_3(q)=\vartheta_{0,2}(q)+\vartheta_{2,2}(q),\qquad\theta_4(q)=\vartheta_{0,2}(q)-\vartheta_{2,2}(q).
\ee

\section{Rewriting of the bulk free energy}
\label{AppBFE}

The function ${\rm arcsinh}(\alpha\sin t)$ appears in the ground-state energy (\ref{gseven}) and in the expression (\ref{fbulk}) for the bulk free energy per site. Representing this function by an integral yields
\be
 \int_0^{\frac{\pi}{2}} {\rm d}t \: {\rm arcsinh}{(\alpha \sin{t})}
 = \int_0^{\frac{\pi}{2}} {\rm d}t \: \int_0^1\; {\rm d}x\; \frac{\alpha \sin t}{\sqrt{1+\alpha^2\,x^2\sin^2t}}\,.
\ee
Substituting
\be
 u=\cos t,\qquad y=\alpha\, x,
\ee
and performing the integration over $u$, then gives
\be
 \int_0^{\frac{\pi}{2}} {\rm d}t \: {\rm arcsinh}{(\alpha \sin{t})}
 = \int_0^\alpha {\rm d}y\;\frac{\arctan y}{y}
 =\frac{\ri}{2}\big({\rm Li}_2(-\ri\alpha)-{\rm Li}_2(\ri\alpha)\big).
\ee
Furthermore, it is recalled that the Legendre chi function $\chi^{}_2$ is a special case of the Lerch transcendent $\Phi$, expressible in terms of dilogarithms ${\rm Li}_2$,
\be
 -\ri\chi^{}_2(\ri\alpha)=\frac{\alpha}{4}\Phi(-\alpha^2,2,\half)=\frac{\ri}{2}\big({\rm Li}_2(-\ri\alpha)-{\rm Li}_2(\ri\alpha)\big)=\sum_{j=0}^\infty(-1)^j\frac{\alpha^{2j+1}}{(2j+1)^2}.
\label{chi2}
\ee



\end{document}